\documentclass[12pt]{article}
\usepackage{amsmath}
\usepackage{amsfonts}        
\usepackage{amssymb}
\usepackage{amsbsy}
\usepackage{mathrsfs}
\usepackage{epsfig}      
\usepackage{color}       

\usepackage[T1]{fontenc} 

\usepackage{graphicx}

\newlength{\TZ}
\DeclareFontFamily{OT1}{pzc}{}
\DeclareFontShape{OT1}{pzc}{m}{it}{<-> s * [1.200] pzcmi7t}{}
\DeclareMathAlphabet{\mathpzc}{OT1}{pzc}{m}{it}
\newcommand{\BEQ}{\begin{equation}}     
\newcommand{\BEA}{\begin{eqnarray}}
\newcommand{\BD}{\begin{displaymath}}
\newcommand{\EEQ}{\end{equation}}       
\newcommand{\EEA}{\end{eqnarray}}
\newcommand{\ED}{\end{displaymath}}

\newcommand{\D}{{\rm d}}                
\newcommand{\II}{{\rm i}}               
\newcommand{\erfc}{{\rm erfc\,}}        
\newcommand{\demi}{\frac{1}{2}}         
\newcommand{\wit}[1]{\widetilde{#1}}    
\newcommand{\wht}[1]{\widehat{#1}}      

\renewcommand{\vec}[1]{\boldsymbol{#1}} 

\catcode`\@=11
\def\numberbysection{\@addtoreset{equation}{section}
        \def\theequation{\thesection.\arabic{equation}}}
\numberbysection                        

\definecolor{gruen}{rgb}{0,0.625,0}       
\definecolor{rot}{rgb}{0.75,0,0}          
\definecolor{blau}{rgb}{0,0,0.75}         
\definecolor{casta}{rgb}{0.45,0.20,0}     
\definecolor{gelb}{rgb}{0.825,0.725,0.0}  

\begin{document}

\begin{titlepage}

\vskip 1.5 cm
\begin{center}
{\LARGE \bf Schr\"odinger-invariance in the voter model}
\end{center}

\vskip 2.0 cm
\centerline{{\bf Malte Henkel}$^{a,b}$ and {\bf Stoimen Stoimenov}$^c$}
\vskip 0.5 cm
\centerline{$^a$Laboratoire de Physique et Chimie Th\'eoriques (CNRS UMR 7019),}
\centerline{Universit\'e de Lorraine Nancy, B.P. 70239, F -- 54506 Vand{\oe}uvre l\`es Nancy Cedex, France}
\vspace{0.5cm}
\centerline{$^b$Centro de F\'{i}sica Te\'{o}rica e Computacional, Universidade de Lisboa,}
\centerline{Campo Grande, P -- 1749-016 Lisboa, Portugal}
\vspace{0.5cm}
\centerline{$^c$ Institute of Nuclear Research and Nuclear Energy, Bulgarian Academy of Sciences,}
\centerline{72 Tsarigradsko chaussee, Blvd., BG -- 1784 Sofia, Bulgaria}
\vspace{0.5cm}

\begin{abstract}
Exact single-time and two-time correlations and the two-time response function are found for the order-parameter in the voter model with nearest-neighbour interactions.
Their explicit dynamical scaling functions are shown to be continuous functions of the space dimension $d>0$.
Their form reproduces the predictions of non-equilibrium representations of the Schr\"odinger algebra for models
with dynamical exponent $\mathpzc{z}=2$ and with the dominant noise-source coming from the heat bath.
Hence the ageing in the voter model is a paradigm for relaxations in non-equilibrium critical dynamics, without detailed balance,
and with the upper critical dimension $d^*=2$.
\end{abstract}
\end{titlepage}

\setcounter{footnote}{0}

\section{Introduction: physical ageing in the voter model} \label{sec:1}

The understanding of the collective behaviour of many-body systems out of equilibrium continues to pose many challenges \cite{Taeu14,Cugl15,Giam16,Bait18,Bait22,Vinc24}.
Since exact solutions are only exceptionally available and numerical simulations may require large resources in computer facilities,
the use of symmetries may provide additional and complementary insight.
A remarkable and often-studied case is {\em physical ageing} \cite{Stru78}, formally characterised by its three defining properties \cite{Henk10}:
(I) slow relaxational dynamics, (II) absence of time-translation-invariance and (III) dynamical scaling.
In classical systems ageing is usually realised by preparing a system in a totally disordered initial state before quenching it instantaneously
either onto a critical point $T=T_c>0$ or else into the disordered phase with temperature $T<T_c$.
When $T=T_c$, one speaks of {\em non-equilibrium critical dynamics} \cite{Godr02}
whereas for $T<T_c$, one is dealing with {\em phase-ordering kinetics} \cite{Bray94a}.
At $T=T_c$, the dynamics is characterised by critical-point fluctuations,
whereas for $T<T_c$ there is a competition of at least two distinct, but equivalent, equilibrium states and the dynamics is driven by the interface tension between them \cite{Bray94a}.
In either case, the system becomes spatially non-homogeneous and decomposes microscopically into clusters
of a time-dependent and growing linear size $\ell=\ell(t)$.
We shall admit that this growth is algebraic at late times, viz. $\ell(t)\sim t^{1/\mathpzc{z}}$, which defines the dynamical exponent $\mathpzc{z}$.
A convenient characterisation uses the time-space-dependent order-parameter $\phi(t,\vec{r})$
{\bf --} in pure magnetic systems identified as the coarse-grained local magnetisation.
For a totally disordered initial state we can admit that on average the initial order-parameter vanishes, such that for all times
$\left\langle \phi(t,\vec{r})\right\rangle=\left\langle \phi(0,\vec{r})\right\rangle=0$.
The study of such systems is centred on analysing the single-time and {\em two-time correlation function} $C$ and the {\em two-time response function} $R$, defined as
\BEQ \label{gl:1}
C(t,s;{r}) = \bigl\langle \phi(t,\vec{r})\phi(s,\vec{0})\bigr\rangle \;\; , \;\;
R(t,s;{r}) = \left. \frac{\delta \bigl\langle \phi(t,\vec{r})\bigr\rangle}{\delta h(s,\vec{0})}\right|_{h=0}
=\bigl\langle \phi(t,\vec{r})\wit{\phi}(s,\vec{0})\bigr\rangle
\EEQ
where the average is both over initial states as well as over thermal histories.
We shall always admit the habitual spatial translation- and rotation-invariances, such that $\vec{r}\mapsto r = |\vec{r}|$, for notational simplicity.
In (\ref{gl:1}) we anticipate from Janssen-de Dominicis theory \cite{Domi76,Jans76} that $R$ can be formally rewritten as a correlator with the so-called
{\em response scaling operator} $\wit{\phi}(t,\vec{r})$, to be often-used in this work.
Setting $t=s$ in (\ref{gl:1}) gives\footnote{Practical means of obtaining the length scale $\ell(t)$ include
solving an equation $C\bigl(t;\ell(t)\bigr)=\mathfrak{c}$ with a constant $0<\mathfrak{c}<1$, or calculating the second moment
$\ell^2(t) = \left.\int_{\mathbb{R}^d}\!\D\vec{r}\: r^2\, C(t;\vec{r})\right/\int_{\mathbb{R}^d}\!\D\vec{r}\:C(t;\vec{r})$.}
the {\em single-time correlator} $C(s;r) := C(s,s;r)$.
Setting $r=0$ in (\ref{gl:1}) produces the {\em auto-correlator} $C(t,s):=C(t,s;{0})$ and the {\em auto-response} $R(t,s):=R(t,s;{0})$.
Throughout, we shall assume model-A-type dynamics without any macroscopic conservation law.

\begin{figure}[tb]
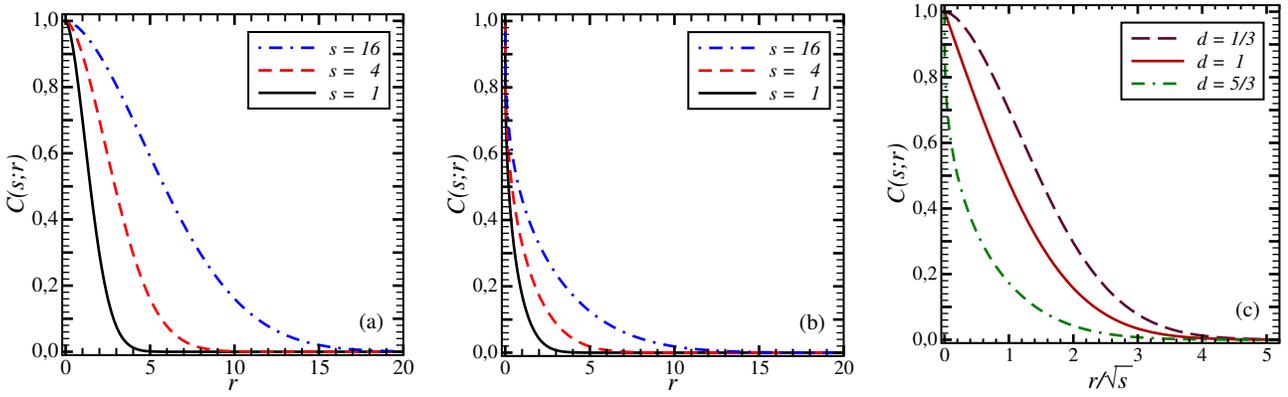

\includegraphics[width=0.32\hsize]{stoimenov17_C1_d0.333_s.eps}  ~
\includegraphics[width=0.32\hsize]{stoimenov17_C1_d1.667_s.eps}  ~
\includegraphics[width=0.31\hsize]{stoimenov17_C1_d_skal.eps}
\caption[fig1]{Physical ageing in the single-time correlator of the voter model for $d<2$ dimensions.
Panel (a) shows $C(s;r)$ for $d=1/3$ and $s=[1,4,16]$.
Panel (b) shows $C(s;r)$ for $d=5/3$ and the same three values of $s$.
Panel (c) shows the associated scaling function of $C(s;r)=F_C\bigl(1;r/\sqrt{s\,}\,\bigr)$ for the dimensions $d=[1/3,1,5/3]$ from top to bottom.
\label{fig1} }
\end{figure}

\begin{figure}[tb]
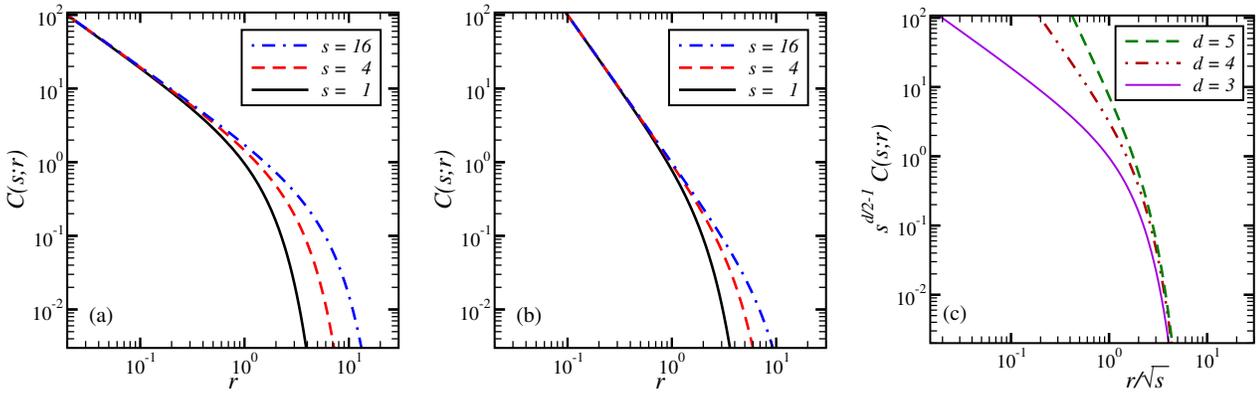

\includegraphics[width=0.31\hsize]{stoimenov17_C1_d3.0_s.eps}  ~
\includegraphics[width=0.31\hsize]{stoimenov17_C1_d4.0_s.eps}  ~
\includegraphics[width=0.31\hsize]{stoimenov17_Cs_d_skal.eps}
\caption[fig2]{Physical ageing in the single-time correlator of the voter model for $d>2$ dimensions.
Panel (a) shows $C(s;r)$ for $d=3$ and $s=[1,4,16]$.
Panel (b) shows $C(s;r)$ for $d=4$ and the same values of $s$.
Panel (c) shows the scaling function $s^{d/2-1} C(s;r)=F_C\bigl(1;r/\sqrt{s\,}\,\bigr)$  for the dimensions $d=[3,4,5]$ from bottom to top.
\label{fig2} }
\end{figure}

In this work, we shall study the physical ageing of the so-called {\em voter-model} \cite{Ligg85,Ligg99,Tome01,Krap10},
in $d>0$ spatial dimensions which we shall consider as a continuous variable.
Precise definitions will be given in section~\ref{sec:2}. Before that, we shall begin with a qualitative overview,
immediately in the continuum limit, first on the generic ageing behaviour and second on the model-specific properties.
As the first example we consider the single-time correlator, for dimensions $d<2$ in figure~\ref{fig1} and for dimensions $d>2$ in figure~\ref{fig2}.
Figure~\ref{fig1}ab displays $C(s;r)$ for two examples where the dimension $d<2$, for several values of the time $s$.
In both cases, we see that with increasing $s$, the decay of the
correlator slows down such that the dynamics becomes more slow when the system has grown more old, which is the slow-dynamics property (I).
Property (II) follows since for different values of $s$ one obtains distinct curves.
Finally, the curves collapse onto a single one when replotted over against $r/\sqrt{s}$ and these are shown together in figure~\ref{fig1}c, where for comparison we also
added the curve corresponding to $d=1$. This demonstrates property (III).
The same exercise is carried out in figure~\ref{fig2} for two examples with $d>2$.
As before, we can verify the first two defining properties (I) and (II) in figure~\ref{fig2}ab. Property (III) is the data collapse with respect to the variable
$r/\sqrt{s\,}$ when the single-time correlator is rescaled by $s^{d/2-1}$. In figure~\ref{fig2}c we display the shape of the scaling function, for three different values of $d$.


\begin{figure}[tb]
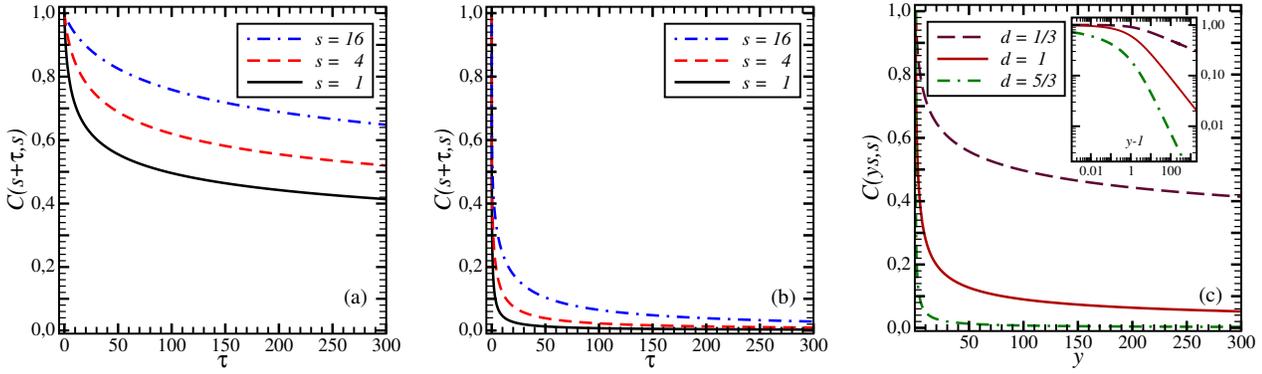

\includegraphics[width=0.31\hsize]{stoimenov17_C2_d0.333_tau.eps}  ~
\includegraphics[width=0.31\hsize]{stoimenov17_C2_d1.667_tau.eps}  ~
\includegraphics[width=0.31\hsize]{stoimenov17_C2_d_skaly_inset.eps}
\caption[fig3]{Physical ageing in the two-time auto-correlator $C(t,s)$ of the voter model for $d<2$ dimensions.
Panel (a) shows $C(s+\tau,s)$ for $d=1/3$ and three values of $s$.
Panel (b) shows $C(s+\tau,s)$ for $d=5/3$, the same values of $s$ and at the same scale.
Panel (c) shows the associated scaling function $C(ys,s)=f_C(y)$ for the dimensions $d=[1/3,1,5/3]$. The inset displays
$f_C(y)$ over against $y-1$, in order to show the power-law decay for $y\gg 1$.
\label{fig3} }
\end{figure}
\begin{figure}[tb]
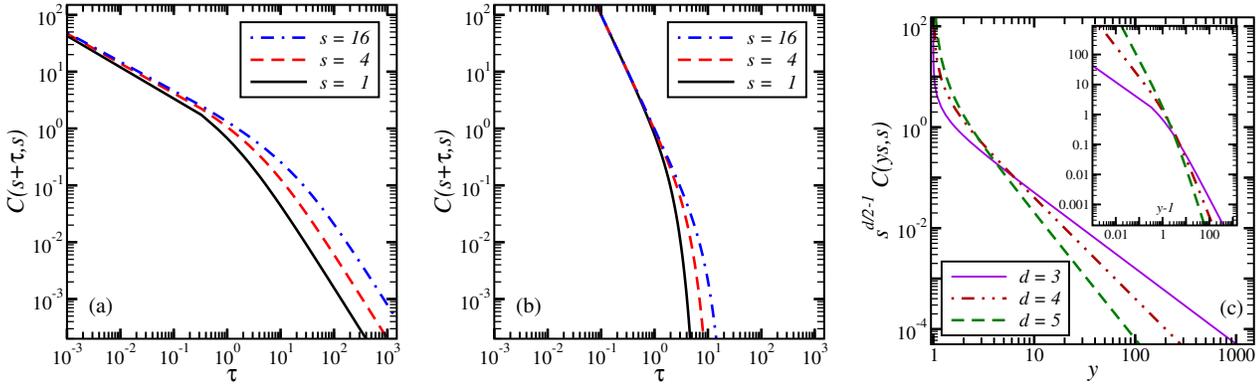

\includegraphics[width=0.31\hsize]{stoimenov17_C2_d3.0_tau.eps}  ~
\includegraphics[width=0.31\hsize]{stoimenov17_C2_d4.0_tau.eps}  ~
\includegraphics[width=0.31\hsize]{stoimenov17_Cs2_d_skal_inset.eps}
\caption[fig4]{Physical ageing in the two-time correlator of the voter model for $d>2$ dimensions.
Panel (a) shows $C(s+\tau,s)$ for $d=3$ and three values of $s$.
Panel (b) shows $C(s+\tau,s)$ for $d=4$, the same values of $s$, at the same scale.
Panel (c) shows the scaling function $s^{d/2-1} C(ys,s)=f_C(y)$ for the dimensions $d=[3,4,5]$. The inset displays
$f_C(y)$ over against $y-1$, and shows the cross-over between two power-laws for $y\gg 1$ and $y-1\ll 1$, respectively.
\label{fig4} }
\end{figure}

Analogously, we now consider the ageing behaviour as it manifests itself in the two-time auto-correlator
$C(t,s)$ which is shown in figure~\ref{fig3} for $d<2$ and
figure~\ref{fig4} for $d>2$. Following the analogy with the single-time correlator,
we show in figure~\ref{fig3}ab the auto-correlator, over against the time difference $\tau=t-s$,
for two examples of dimensions $d<2$, for three values of the {\em waiting time} $s$.
As before, we can verify the three defining properties of ageing: (I) with increasing values of $s$, the auto-correlator decays more slowly,
(II) for different values of $s$ one has distinct curves such that $C(t,s)=C(s+\tau,s)$ does not only depend on $\tau$ and
(III) these curves collapse onto a single one when replotted over against $y=t/s$,
see figure~\ref{fig3}c, where again we also added the curve with $d=1$ for later comparison.
The same exercise is carried out in figure~\ref{fig4} for dimensions $d>2$, with analogous results.
The properties (I) and (II) are checked in figure~\ref{fig4}ab and the shape of the scaling function is shown in figure~\ref{fig4}c,
for the same values of $d$ as above for the single-time correlator, which demonstrates property (III).
Overall, we can conclude that the voter model in all dimensions $d>0$
does satisfy the generic properties of physical ageing.\footnote{This conclusion also holds true for $d=2$, but with 
logarithmic corrections to the scaling behaviour. See sections~\ref{sec:2} and~\ref{sec:4} below.}
Formally, this can be cast into the scaling forms
\begin{subequations} \label{gl:local}
\BEQ \label{gl:2}
C(t,s;{r}) = s^{-b}   F_C\left( \frac{t}{s}; \frac{\bigl|\vec{r}\bigr|}{s^{1/\mathpzc{z}}}\right) \;\; , \;\;
R(t,s;{r}) = s^{-1-a} F_R\left( \frac{t}{s}; \frac{\bigl|\vec{r}\bigr|}{s^{1/\mathpzc{z}}}\right)
\EEQ
with the {\em ageing exponent}s $b,a$. For non-equilibrium critical dynamics at $T=T_c$ in pure magnetic systems, one generically expects that
$a=b=(d-2+\eta)/\mathpzc{z}$ \cite{Godr02,Cala05,Henk10,Taeu14}, where $\eta$ is a standard equilibrium critical exponent;
whereas for phase-ordering kinetics at $T<T_c$ in magnetic
systems one expects $b=0$ but $a=\demi$ and $\mathpzc{z}=2$, at least in pure systems with short-ranged interactions \cite{Bray94a,Henk04PP,Henk10}.
In addition, one usually finds asymptotically for $y\gg 1$
\BEQ \label{gl:3a}
f_C(y) = F_C(y;0) \sim y^{-\lambda_C/\mathpzc{z}} \;\; , \;\; f_R(y) = F_R(y;0) \sim y^{-\lambda_R/\mathpzc{z}}
\EEQ
where $\lambda_C$ is the {\em auto-correlation exponent} \cite{Huse89} and $\lambda_R$ is the {\em auto-response exponent} \cite{Pico02}.
This algebraic form is bourne out in figures~\ref{fig3}c and~\ref{fig4}c in the voter model for the auto-correlator, for all values of the dimension $d$.
In almost all models, provided only that the initial correlations are spatially short-ranged, one also expects that \cite{Bray91,Bray94a,Godr02,Pico04}
\BEQ \label{lambda}
\lambda = \lambda_C = \lambda_R
\EEQ
\end{subequations}
The scaling functions $F_{C,R}\bigl(y,r s^{-1/\mathpzc{z}}\bigr)$ in (\ref{gl:2}) are expected {\em universal},
by which it is meant that their form should be independent of microscopic `details', such as the lattice structure,
the precise form of the interactions, or of temperature upon a quench into $T<T_c$.  On the other hand, one expects them to be $d$-dependent.

Having established that the generic set-up of physical ageing is applicable to the nearest-neighbour voter model,
we now consider its more specific aspects which involve the form of the scaling functions.
In all cases, one has the dynamical exponent $\mathpzc{z}=2$.
For the single-time correlator $C(s;r)$, comparison of their shapes in figures~\ref{fig1}c and~\ref{fig2}c, respectively, shows that they are very different.
Although in all dimensions, the decay for large values of the scaling variable $r/\sqrt{s\,}$
is more fast than any power-law, for small values of this scaling variable there is
a saturation at $C(s;0)=1$ for $d<2$ but a power-law behaviour in $r/\sqrt{s\,}\ll 1$ whenever $d>2$. In figure~\ref{fig1}c we see that when $d<2$
the $r$-dependent approach towards $C(s;0)=1$ depends
clearly on the value of $d$.\footnote{In phase-ordering kinetics of pure magnets, at $T<T_c$, the celebrate Porod's law stipulates that
$C(s;\vec{r})\simeq 1 - \mbox{\rm cste.} \frac{|\vec{r}|}{\sqrt{s\,}}+\ldots$ for $|\vec{r}|/\sqrt{s\,}\ll 1$ \cite{Poro51,Bray94a}.
In the voter model, this only holds for $d=1$, see figure~\ref{fig1}c.} When $d>2$, the respective exponent is seen to be $d$-dependent from figure~\ref{fig2}c.
We also notice that while for $d<2$ the single-time correlator scales directly,
for $d>2$ the single-time correlator must be rescaled according to $s^{d/2-1}C(s;r)$ in order to achieve the scaling collapse
and that the scaling functions decrease with $d$ for $d<2$ while they increase with $d$ for $d>2$ (in our normalisation, see section~\ref{sec:2}).
The reason behind such different behaviour calls for an explanation.
Similarly, we compare the shapes of the two-time auto-correlator $C(t,s)$ in figures~\ref{fig3}c and~\ref{fig4}c.
Again for $d<2$ and $d>2$, respectively, we find very different shapes and possibly distinct dependencies on $d$.
Although for large values of the scaling variable $y=t/s$ the decay is algebraic for all dimensions $d>0$
(which also shows that the relaxation times are formally infinite such that the ageing system will never reach a stationary state) and in agreement with the expected (\ref{gl:3a}),
we also observe here a saturation at $C(s,s)=1$ for $d<2$ with a $d$-dependent approach as a function of $y-1$
(inset of figure~\ref{fig3}c)  while there
is a power-law behaviour in $y-1\ll 1$ for $d>2$, with a $d$-dependent exponent (inset of figure~\ref{fig4}c).
Also, we observe that for $d<2$, the auto-correlator $C(ys,s)$ is $s$-independent and directly scales
whereas for $d>2$, a data collapse is seen for the rescaling auto-correlator $s^{d/2-1}C(ys,s)$. An explanation for this must be sought as well.
This will come from the exact expressions for all these scaling functions which we shall calculate explicitly.

Finally, we shall use the exactly solvable voter model to test generic ideas on the form of these scaling functions, which is our main motivation to undertake this work.
Evidently, all observations for which we provided examples in the figures will be confirmed from the exact results, for all $d>0$.
{}From the details of its definition (see section~\ref{sec:2}) it will
become apparent that the voter model is actually at a critical point between two non-critical phases \cite{Krap92}. Then the scaling form of its observables will be determined
by the fluctuations coming from the thermal bath but will be independent of the fluctuations of the initial state (in the renormalisation-group sense, the latter ones are irrelevant and will
contribute at most to corrections to the leading behaviour, onto which we shall focus exclusively).
Since the dynamical exponent $\mathpzc{z}=2$, it is tempting to generalise the dynamical scaling symmetry, by definition
present in any physically ageing system, and demonstrated to hold for the voter model in figures~\ref{fig1}c,\ref{fig2}c,\ref{fig3}c and~\ref{fig4}c, to a larger symmetry. The natural
candidate for this is the Schr\"odinger group, which is known to be a dynamical symmetry of the free diffusion equation from the work of Jacobi and Lie in the middle of the
19$^{\rm th}$ century, see \cite{Duva24} for a historical review. The first applications to dynamics in statistical physics date at least back to \cite{Henk94}. More recently, it has been
understood how to adapt the (equilibrium) predictions of Schr\"odinger-invariance, by a change of the Lie algebra representation, to the far-from-equilibrium context of physical
ageing \cite{Henk25,Henk25c}.\footnote{This seems to be unrelated to the `two-time physics' studied in quantum gravity and which involves a two-dimensional time \cite{Bars01}.} 
Especially for systems with a dynamical exponent $\mathpzc{z}=2$
and which undergo non-equilibrium critical dynamics for which the fluctuations of the heat bath are relevant,
very recently we derived the generic form of the scaling function $F_C\bigl(1;r/\sqrt{s\,}\,\bigr)$ of the single-time correlator and also for
$f_C(y)$ of the two-time auto-correlator ({\it modulo} a technical hypothesis which will turn out to be valid in the voter model) \cite{Henk25d}, see section~\ref{sec:3}.\footnote{Previous
attempts of testing Schr\"odinger-invariance studied only the response function $R(t,s;r)$.}
By construction, the results of a dynamical symmetry, if applicable, are universal.
In addition, the dynamical Schr\"odinger symmetry reduces the problem of finding the form of a scaling function (which in principle requires to determine infinitely many parameters) to the
determination of a finite (and small) set of parameters and whose values are furthermore related between different observables.
We shall use the voter model, in any dimension $d>0$, to test (and eventually confirm) these predictions.
It is suggestive to consider non-integer values of $d$ as akin to study the voter model on a fractal \cite{Such06,Bab08,Godr24} although it is known
that a fractal geometry is characterised by more than one fractal dimension \cite{Ramm83,Pati24,Lima24,Fume25}.\footnote{The relationship between the geometric fractal (Hausdorff)
dimension $d_f$ and the dynamical scaling dimension $d_{\rm eff}$ seems also to depend on the fractal,
see \cite{Genz23,Genz25,Alme25} for recent studies on equilibrium phase transitions in the Ising model and references therein.}
We shall also include an explicit treatment when the voter model is at its upper critical dimension $d=d^*=2$.
We shall see that a different non-equilibrium representation of the Schr\"odinger Lie algebra must be used.
Altogether, this work provides a new kind of evidence that the $d$-dimensional voter model falls into the class of non-equilibrium critical dynamics.

This work is organised as follows. In section~\ref{sec:2} we present the exact solution of the voter model, with nearest-neighbour interactions,
and derive explicit expressions for the correlators $C(s;r)$ and $C(t,s)$ as well as for the response function $R(t,s;r)$ whose scaling functions depend continuously on $d>0$.
In section~\ref{sec:3} we recall the necessary background on non-equilibrium field-theory, on Schr\"odinger-invariance at equilibrium
and on how to map the results onto non-equilibrium observables. In section~\ref{sec:4} we compare the
exact results from section~\ref{sec:2} with the field-theoretic predictions of non-equilibrium Schr\"odinger-invariance and thereby
confirm 
that the non-equilibrium voter model is indeed Schr\"odinger-invariant for all dimensions $d>0$. We conclude in section~\ref{sec:5}.

\section{The voter model} \label{sec:2}

The much-studied voter model can be formulated as a classical spin model, defined in terms of spin variables $\sigma_{\vec{n}}=\pm 1$ attached to the sites
$\vec{n}\in\Lambda\subset\mathbb{Z}^d$ of a hyper-cubic lattice in $d$ dimensions.
A spin configuration is denoted as $\{\sigma\} = \bigl(\sigma_1,\ldots,\sigma_{\cal N}\bigr)$,
where ${\cal N}=|\Lambda|$ is the total number of sites. That configuration arises with the probability $P\bigl(\{\sigma\},t\bigr)$.
The dynamics is described in terms of a master equation
\BEQ
\partial_t P\bigl( \{\sigma\},t) = \sum_{\{\sigma'\}} \left[ w\bigl(\{\sigma' \}\to \{\sigma \}\bigr) P\bigl(\{\sigma'\},t)
-w\bigl(\{\sigma \}\to \{\sigma' \}\bigr) P\bigl(\{\sigma\},t) \right]
\EEQ
In the voter model, transitions between configurations occur via single spin flips.
If the spin to be flipped is at site $\vec{n}$, the transition rates of the {\em voter model} are \cite{Ligg85,Ligg99,Tome01,Krap10}
\BEQ \label{gl:2.2}
w\bigl(\{\sigma \}\to \{\sigma' \}\bigr) ~~\mapsto~~
w_{\vec{n}}\bigl(\{\sigma \}\bigr) =  \demi\left( 1 - \frac{1}{2d} \sigma_{\vec{n}} \sum_{\vec{m}(\vec{n})} \sigma_{\vec{m}} \right)
\EEQ
where $\vec{m}(\vec{n})$ are the nearest-neighbour sites with respect to the site $\vec{n}\in\Lambda$.
The model appears to have been first introduced in the 1960s in studies of genetic
correlations before receiving profound interest from probability theory, see \cite{Ligg85,Corb24e} and refs. therein.
It has numerous applications: for example it is often construed (possibly with generalisations) as a simple model for opinion-forming
\cite{Cast09,Redn09,Fern14} or is used in the modelling of network dynamics \cite{Vasq08,Carr16,Doro22,Bern23}.
In a physical context, it was re-discovered either as a simple model of a non-equilibrium spin system subject to two distinct baths,
each creating its proper dynamics at its own temperature \cite{Droz89,Droz90}, or in surface catalytic reactions \cite{Krap92,Oliv03}\footnote{The specific formulation
of the model makes it clear that it sits on a critical line between two non-critical absorbing phases.}
or as a prototype of critical-point ageing in models with several absorbing states \cite{Dorn01}.
It is one of the very few models of interacting spins which is analytically solvable in any number of dimensions, for any kind of interactions \cite{Krap92,Frac96,Frac97,Vasq08}.
With the widely studied Glauber-Ising model \cite{Glau63} it shares the invariance under a total
spin-reversal $\sigma_{\vec{n}}\mapsto -\sigma_{\vec{n}}$ for all $\vec{n}\in\Lambda$.
In contrast to the Glauber-Ising model, the voter model does not satisfy the detailed-balance condition
\cite{Ligg85,Ligg99,Tome01,Krap10,Godr13}\footnote{With the only exception of the case $d=1$ when the voter model reduces to the Glauber-Ising chain and then
of course does obtain detailed balance \cite{Glau63}.}
such that its stationary states cannot be at thermal equilibrium. Rather, the voter model has two absorbing states,
namely $\sigma_{\vec{n}}=1$ for all $\vec{n}\in\Lambda$ (or $\sigma_{\vec{n}}=-1$ for all $\vec{n}$) into which the system
may enter but it cannot leave.
Still, although ``the voter model belongs to a different universality class, some of [its]
features \ldots have a clear counterpart in ferromagnetic models'' \cite{Corb24e}. The solubility of the voter model has been used to
carry out a detailed set of studies on its ageing behaviour in the presence of long-range interactions with distant-dependent interaction rates
$w(r)\sim r^{-\alpha}$, first numerically for $d=1$ \cite{Rodr11}
and recently in a long series of analytical studies \cite{Corb24a,Corb24b,Corb24c,Corb24d,Corb24e,Corb24f} in $d=1,2,3$ dimensions. While for $\alpha >2+d$, the late-time behaviour
is the same as for the nearest neighbour rates (\ref{gl:2.2}), a very rich and new behaviour is found for smaller values of $\alpha$. Since in these
new long-ranged regimes, the dynamical exponent $\mathpzc{z}<2$ if it exists at all \cite{Corb24a,Corb24b,Corb24e},
we cannot use those results in our planned comparison with Schr\"odinger-invariance
and therefore shall restrict ourselves to the nearest-neighbour rates (\ref{gl:2.2}).
However, we shall consider the continuous dependence of the scaling functions on the dimension $d>0$
which allows to take advantage of the voter model as a paradigmatic case of ageing without detailed balance.
New insight beyond what would be accessible when restricting to $d=1,2,3$ will be obtained.

Before we begin our analysis of correlators and responses, we add two more comments. First, consider the average order parameter
$S_{\vec{n}}(t) = \bigl\langle \sigma_{\vec{n}}\bigr\rangle(t)$. From the master equation, equations of motion for $S_{\vec{n}}$ are readily derived, using
$\partial_t\langle \sigma_{\vec{n}}\rangle = -2 \langle \sigma_{\vec{n}} w_{\vec{n}}\rangle$, and
lead to \cite{Krap92} $S_{\vec{n}}(t) = e^{-dt} \sum_{\vec{m}} S_{\vec{n}-\vec{m}}(0) I_{n_1-m_1}(t)\cdots  I_{n_d-m_d}(t)$. Since for a fully disordered
initial state $S_{\vec{n}}(t) = S_{\vec{n}}(0)=0$ for all times $t>0$,
at least on average {the two stationary states are never reached in a spatially infinite system}.
Second, the density $n_{\rm r}(t)$ of reactive $`+-'$-interfaces, expressed in terms of a nearest-neighbour correlator,
is evaluated for large times as \cite{Frac96,Oliv03,Cast09}
\BEQ
n_{\rm r}(t) = \demi \biggl( 1 - \bigl\langle \sigma_{\vec{n}} \sigma_{\vec{n}+\vec{1}} \bigr\rangle \biggr) \sim
\left\{ \begin{array}{ll} t^{-(2-d)/2}                         & \mbox{\rm ~~;~ if $d<2$} \\
                          1/\ln t                              & \mbox{\rm ~~;~ if $d=2$} \\
                          \mathfrak{a} - \mathfrak{b} t^{-d/2} & \mbox{\rm ~~;~ if $d>2$}
        \end{array} \right.
\EEQ
where $\mathfrak{a},\mathfrak{b}$ are constants. This points to an important difference between the $d<2$ and $d>2$ cases:
it is generally said that for $d\leq 2$ the voter model undergoes {\em coarsening} since $n_{\rm r}(t)\to 0$ and for $d>2$,
{there are infinitely many stationary states} since $n_{\rm r}(t)$ remains finite \cite{Ligg85,Ligg99}.

\subsection{Single-time correlator}

In what follows, the spatial continuum limit will be taken throughout.
Then the single-time spin-spin correlator $C_{\vec{n}}(t)=\bigl\langle \sigma_{\vec{n}} \sigma_{\vec{0}}\bigr\rangle(t)$
will become $C(t;\vec{r})=\bigl\langle \phi(t,\vec{r})\phi(t,\vec{0})\bigr\rangle$ in terms of a coarse-grained order-parameter $\phi(t,\vec{r})$. Inserting
the rate (\ref{gl:2.2}) into the master equation leads after rescaling to the equation of motion for the single-time correlator
\BEQ \label{gl:2.4}
\partial_t C(t;r) = \Delta_{\vec{r}} C(t;r) = \left( \partial_r^2 + (d-1)\frac{1}{r}\partial_r \right) C(t;r)
\EEQ
where $\Delta_{\vec{r}}$ is the spatial laplacian and we also used that in the continuum limit, spatial translation- and rotation-invariance are restored.
Then the correlator merely depends on the absolute value $r=|\vec{r}|$.
Eq.~(\ref{gl:2.4}) will be the basis of all subsequent calculations.
To solve this equation,\footnote{For tutorials on solving explicitly (fractional) partial differential equations, see \cite{Mana24,Kulm25}.}
we use the scaling ansatz \cite{Krap92,Tome01,Corb24e}
\BEQ \label{gl:2.5}
C(t;r) = t^{-b} F_C\bigl( \mathfrak{u} \bigr) \;\; , \;\; \mathfrak{u} = r t^{-1/2}
\EEQ
where $b$ will be identified later on with one of the ageing exponents. Clearly, $\mathpzc{z}=2$.
Eq.~(\ref{gl:2.5}) implies the double scaling limit $t\to\infty$, $r\to\infty$ such that $\mathfrak{u}$ is kept fixed.
{}From now on $d$ can be treated as a continuous parameter. It follows that the scaling function obeys the differential equation
\BEQ \label{gl:2.6}
F_C''(\mathfrak{u}) + (d-1)\frac{1}{\mathfrak{u}}F_C'(\mathfrak{u}) +\demi \mathfrak{u} F_C'(\mathfrak{u}) + b F_C(\mathfrak{u}) = 0
\EEQ
with the general solution \cite{Kamk79}
\BEQ \label{gl:gen-solC1}
F_C(\mathfrak{u}) = \mathscr{C}_1\, e^{-\mathfrak{u}^2/4}\, M\left( \frac{d}{2}-b,\frac{d}{2}; \frac{\mathfrak{u}^2}{4} \right)
+ \mathscr{C}_2\, e^{-\mathfrak{u}^2/4}\, U\left( \frac{d}{2}-b,\frac{d}{2}; \frac{\mathfrak{u}^2}{4} \right)
\EEQ
where $M,U$ are the confluent Kummer hypergeometric functions \cite{Abra65} and $\mathscr{C}_{1,2}$ are constants.

Next, we find the value of the exponent $b$ from the stationary solution which obeys \cite{Tome01}
\BEQ
\Delta_{\vec{r}} C(\infty;r) = 0 ~~\Longrightarrow~~ \left( \partial_{\mathfrak{u}}^2 + (d-1)\frac{1}{\mathfrak{u}}\partial_{\mathfrak{u}}\right) f(\mathfrak{u}) =0
\EEQ
with the solution $f(\mathfrak{u}) = f_0 + f_1 \mathfrak{u}^{2-d}$ where $f_{0,1}$ are constants. Because of the definition (\ref{gl:2.5}) of
$\mathfrak{u}$, the stationary limit $t\to\infty$ corresponds to $\mathfrak{u}\ll 1$. In that limit, we have for $d<2$ that $f(\mathfrak{u})\to f_0$
saturates and for $d>2$ that $f(\mathfrak{u})\to f_1 \mathfrak{u}^{2-d}$ is reminiscent of an equilibrium  critical correlator
$C_{\rm eq}(r)\sim r^{-(d-2+\eta)}$. Combining this with the scaling from (\ref{gl:2.5}) allows us to conclude
\BEQ \label{gl:b-exp}
b = \left\{ \begin{array}{ll}  0       & \mbox{\rm ~~;~ if $d<2$} \\
                               (d-2)/2 & \mbox{\rm ~~;~ if $d>2$}
            \end{array} \right.
\EEQ
The case $d=2$ is treated separately since the ansatz (\ref{gl:2.5}) will no longer work, see section~\ref{subsec:2.4}.

{\bf 1.} Inserting the result (\ref{gl:b-exp}) into (\ref{gl:gen-solC1}) gives for $d<2$
\BEQ
C(t;r) = \mathscr{C}_1\, e^{-r^2/4t}\, M\left( \frac{d}{2},\frac{d}{2}; \frac{r^2}{4t} \right)
+ \mathscr{C}_2\, e^{-r^2/4t}\, U\left( \frac{d}{2},\frac{d}{2}; \frac{r^2}{4t} \right)
= \mathscr{C}_1 + \mathscr{C}_2\, e^{-r^2/4t}\, U\left( \frac{d}{2},\frac{d}{2}; \frac{r^2}{4t} \right)
\EEQ
Since $b=0$, the boundary conditions are that $C(t;r)\to 0$ as $r\to\infty$ more fast than any power and that $C(t;r)\to 1$ as $r\to 0$. Using the known
asymptotics of the Kummer/Tricomi function $U$ \cite{Abra65}, we have for $r\to\infty$ that
$C(t;r)\simeq \mathscr{C}_1 + \mathscr{C}_2\, e^{-r^2/4t}\, \bigl( \frac{r^2}{4t}\bigr)^{-d/2}$ which
can only decay for $\mathscr{C}_1=0$. The second boundary condition leads via \cite[(13.1.6)]{Abra65} to
$\mathscr{C}_2\,\Gamma\bigl(1-\frac{d}{2}\bigr)\stackrel{!}{=}1$, so that
the final solution reads
\begin{subequations} \label{gl:voter-C1}   
\BEQ \label{gl:voter-C1-bas}               
C(t;r) = \frac{1}{\Gamma\bigl(1-\frac{d}{2}\bigr)}\,e^{-r^2/4t}\, U\left(\frac{d}{2},\frac{d}{2};\frac{r^2}{4t} \right)
= \frac{\Gamma\bigl(1-\frac{d}{2},\frac{r^2}{4t}\bigr)}{\Gamma\bigl(1-\frac{d}{2}\bigr)} \;\; ; \;\; d<2
\EEQ
\end{subequations}                        
but can also be expressed in terms of an incomplete Gamma function \cite[(13.6.28)]{Abra65}. It does agree with the generic scaling expectation (\ref{gl:2}).

If $d=1$, the voter model becomes identical to the Glauber-Ising chain and we reproduce the known result
$C^{(1D)}(t;r) = \frac{1}{\sqrt{\pi\,}} e^{-r^2/4t}\,U\bigl(\demi,\demi;\frac{r^2}{4t}\bigr) = \erfc\bigl(\frac{r}{2\sqrt{t\,}}\bigr)$ \cite{Bray97,Godr00a,Corb24a,Henk25b} as expected.

In ferromagnets which undergo phase-ordering kinetics, the single-time correlator satisfies to well-known Porod's law \cite{Poro51,Bray94a}.
In the $d=1$ special case, equivalent to the Glauber-Ising chain, we have the initially linear decay
$C^{(1D)}(t;r) \simeq 1 - \frac{1}{\sqrt{\pi\,}}\frac{|r|}{\sqrt{t\,}}+\ldots$ and Porod's law holds true.
More generally in the voter model, a small-distance expansion of (\ref{gl:voter-C1-bas}) gives
\BEA
C(t;r) &=&  \frac{\Gamma\bigl(\frac{d}{2}-1\bigr)}{\Gamma\bigl(\frac{d}{2}\bigr)\Gamma\bigl(1-\frac{d}{2}\bigr)}
\left( \frac{\vec{r}^2}{4t}\right)^{1-d/2} e^{-\vec{r}^2/4t} {}_1F_1\left(1;2-\frac{d}{2};\frac{\vec{r}^2}{4t}\right)
+ e^{-\vec{r}^2/4t} {}_1F_1\left(\frac{d}{2};\frac{d}{2};\frac{\vec{r}^2}{4t}\right)
\nonumber \\
&\simeq& 1 - \frac{2^{d-2}}{\Gamma\bigl(2-\frac{d}{2}\bigr)} \left( \frac{\vec{r}^2}{t} \right)^{1-d/2} + \ldots \;\; ; \;\; d<2
\EEA
which does not lead to a linear decay of the single-time correlator for small $r$ and Porod's law does not hold for $d\ne 1$, see also figure~\ref{fig1}c.
Since for small $d$, the initial decay is more slow than linear and for dimensions close to $d=2$,
it is much more fast, by continuity we would have expected to find an intermediate
dimension, where the decay happens to be linear, as it occurs indeed for $d=1$.
In addition, we can also consider the structure factor
\BEA
\wht{S}(t;\vec{q}) &:=& \int_{\mathbb{R}^d}\!\D\vec{r}\: e^{-\II\vec{q}\cdot\vec{r}}\, C(t;r) \:=\:
\frac{1}{\Gamma\bigl(1-\frac{d}{2}\bigr)} \int_{\mathbb{R}^d} \!\D\vec{r}\: e^{-\II\vec{q}\cdot\vec{r} -\frac{1}{4}\vec{r}^2/t}\,
U\left(\frac{d}{2},\frac{d}{2};\frac{\vec{r}^2}{4t}\right)
\nonumber \\
&=& \frac{\bigl(4\pi\bigr)^{d/2} \Gamma\bigl(\frac{d}{2}+1\bigr)}{\Gamma\bigl(1-\frac{d}{2}\bigr)\Gamma\bigl(\frac{d}{2}\bigr)} t^{d/2}\,
{}_1F_1\left(1;\frac{d}{2}+1;-t\vec{q}^2\right) \;\; ; \;\; d<2
\EEA
which was evaluated via the integral representation \cite[(13.2.5)]{Abra65} for $U$.
This certainly has the expected scaling form $\wht{S}(t;\vec{q})=\wht{S}_0 q^{-d} f\bigl(\vec{q}^2 t\bigr)$.
But asymptotically for $q=|\vec{q}|\to\infty$, we have with \cite[(13.5.1)]{Abra65} that $\wht{S}(t;\vec{q})\sim t^{d/2-1} q^{-2}$.
This agrees with the expected Porod form $\sim q^{-d-1}$ \cite{Poro51,Bray94a} only for $d=1$.
We conclude that unsurprisingly, the dynamics of the voter model is in general different from the phase-ordering kinetics of a ferromagnet at $T<T_c$,
since generically, it does not satisfy detailed balance.

{\bf 2.} For $d>2$, we have from (\ref{gl:b-exp}) that $b=(d-2)/2$ and read off from (\ref{gl:gen-solC1})
\BEQ
C(t;r) = \mathscr{C}_1\, t^{-(d-2)/2}\, e^{-r^2/4t}\, M\left( 1,\frac{d}{2}; \frac{r^2}{4t} \right)
+ \mathscr{C}_2\, t^{-(d-2)/2}\,e^{-r^2/4t}\, U\left( 1,\frac{d}{2}; \frac{r^2}{4t} \right)
\EEQ
For $r\to\infty$, with \cite[(13.1.4)]{Abra65} and \cite[(13.1.8)]{Abra65} this gives
\BEQ
C(t;r) \simeq \mathscr{C}_1\,\Gamma\left(\frac{d}{2}\right)\,r^{2-d} + \mathscr{C}_2\, t^{-d/2}\, r^{-2}\, e^{-r^2/4t}
\EEQ
which should vanish more fast with $r$ than any power. This is only possible for $\mathscr{C}_1=0$. On the other hand, for $r\to 0$ we find \cite[(13.5.8)]{Abra65}
$C(t;r)\simeq \mathscr{C}_2\,2^{d-2}\Gamma\bigl(\frac{d}{2}-1\bigr)\,r^{2-d}$. Here, since we have taken the continuum limit from the outset, it is no longer possible to achieve a
saturation. One might arbitrarily prescribe a fixed value $C(t;\mathfrak{a})$ at some `small' distance $\mathfrak{a}$, i.e. \cite{Krap92,Frac96,Krap10,Tart15,Corb24e},
but we rather prefer to accept that in the continuum limit, the stationary scaling solution $C(\infty;r)\sim r^{-(d-2)}$ does not saturate when $r\to 0$.\footnote{Alternatively, we might
refrain from taking the continuum limit at all and work out the exact solution $C_{\vec{n}}(t)$ on a discrete lattice, subject to the boundary condition $C_{\vec{0}}(s)=1$ which
would fix $\mathfrak{C}_0$, see \cite{Frac96,Tart15}. But this will deliberately obscure the continuum limit of the scaling functions we are interested in.} Then at most we can impose an
equality of amplitudes, of the kind $\mathscr{C}_2\,\Gamma\bigl(\frac{d}{2}-1\bigr)2^{d-2}\stackrel{!}{=} \mathfrak{C}_0$ such that our final solution is
\addtocounter{equation}{-5}   
\begin{subequations}
\addtocounter{equation}{1}    
\BEQ \label{gl:voter-C1-haut} 
C(t;r) = \frac{\mathfrak{C}_0}{2^{d-2}\Gamma\bigl(\frac{d}{2}-1\bigr)} t^{-d/2+1}\,e^{-r^2/4t}\, U\left(1,\frac{d}{2};\frac{r^2}{4t}\right)
= \mathfrak{C}_0\, r^{2-d} \frac{\Gamma\bigl(\frac{d}{2}-1,\frac{r^2}{4t}\bigr)}{\Gamma\bigl(\frac{d}{2}-1\bigr)} \;\; ; \;\; d>2
\EEQ
\end{subequations}
\addtocounter{equation}{4}   
\noindent
and $\mathfrak{C}_0$ is the amplitude of the stationary correlator for $d>2$. The two equations (\ref{gl:voter-C1}) together give the single-time correlator for all dimensions $d\ne 2$ and
will serve as input for the calculations of the other observables. They prove analytically what we showed by example in figures~\ref{fig1}c and~\ref{fig2}c.
This suggests that for $d<2$ the voter model still shares some properties with systems undergoing phase-ordering kinetics,
whereas for $d>2$ it behaves more like a system undergoing non-equilibrium critical dynamics.

If $d=3$, we have with \cite[(13.1.7)]{Abra65}
\BEA
C^{(3D)}(t;r) &=& \frac{\mathfrak{C}_0}{2\Gamma\bigl(\demi\bigr)} t^{-1/2}\,e^{-r^2/4t}\, U\left(1,\frac{3}{2};\frac{r^2}{4t}\right) \nonumber \\
&=& \frac{\mathfrak{C}_0}{2\sqrt{\pi\,}} t^{-1/2}\,e^{-r^2/4t} \left( \frac{r^2}{4t}\right)^{-1/2}\,U\left(\demi,\demi;\frac{r^2}{4t}\right)
\:=\: \mathfrak{C}_0\,r^{-1} \erfc\left(\frac{r}{2\sqrt{t\,}}\right)
\EEA
which via \cite[(13.6.39)]{Abra65} does reproduce the known result \cite[eq.~(30)]{Corb24e}, up to the choice of the scaling amplitude.

\subsection{Two-time auto-correlator}

Applying again the rates (\ref{gl:2.2}), the two-time correlator $C(t,s;r)$ is found by solving the differential equation
\BEQ \label{gl:2tC}
\partial_t C(t,s;\vec{r}) = \demi \Delta_{\vec{r}} C(t,s;\vec{r}) \;\; , \;\; C(s,s;\vec{r}) = C(s;\vec{r})
\EEQ
together with an initial condition in the equal-time case.
Fourier-transforming $C(t,s;\vec{r})$ with respect to the spatial coordinate $\vec{r}$ straightforwardly leads to the auto-correlator, again with $d$ as a continuous parameter
\BEA
C(t,s;\vec{0}) &=& \frac{1}{(2\pi)^d} \int_{\mathbb{R}^d} \!\D\vec{q}\: e^{-\demi\vec{q}^2(t-s)} \int_{\mathbb{R}^d} \!\D\vec{r}\:
\mathscr{C}_2\,s^{-b}\, e^{-\II\vec{q}\cdot\vec{r} -\vec{r}^2/4s}\, U\left(\frac{d}{2}-b,\frac{d}{2};\frac{\vec{r}^2}{4s}\right)
\nonumber \\
&=& \frac{\mathscr{C}_2}{(2\pi)^d} s^{-b} \int_{\mathbb{R}^d} \!\D\vec{r}\: U\left( \frac{d}{2}-b,\frac{d}{2};\frac{\vec{r}^2}{4s}\right)
\exp\left[-\vec{r}^2\left( \frac{1}{4s} +\frac{1}{2(t-s)}\right)\right] \left( \frac{\pi}{(t-s)/2}\right)^{d/2}
\nonumber \\
&=& \frac{2^d\mathscr{C}_2}{(2\pi)^{d/2}} s^{-b} \left(\frac{t}{s}-1\right)^{-d/2} \int_{\mathbb{R}^d} \!\D\vec{w}\:
e^{-\vec{w}^2\bigl(1+\frac{2}{t/s-1}\bigr)} U\left( \frac{d}{2}-b,\frac{d}{2}; \vec{w}^2\right)
\label{gl:2.16}
\EEA
where we carried out the $\vec{q}$-integration and then changed integration variables.
The scaling collapse follows from the unique dependence on $y=t/s$, up to the $s^{-b}$-prefactor.
For $y=t/s\gg 1$ we can already confirm by comparing with (\ref{gl:3a}) that the auto-correlator exponent $\lambda=d$ for all $d\ne 2$.

{\bf 1.} In the first case $d<2$, we have $b=0$. Using the integral representation \cite[(13.2.5)]{Abra65}, we have from (\ref{gl:2.16}) for $y>1$
\BEA
C(ys,s) &=& \frac{(2/\pi)^{d/2}}{\Gamma\bigl(1-\frac{d}{2}\bigr)\Gamma\bigl(\frac{d}{2}\bigr)} \bigl(y-1\bigr)^{-d/2}
\int_{\mathbb{R}^d} \!\D\vec{w}\: e^{-\vec{w}^2\frac{y+1}{y-1}} \int_0^{\infty} \!\D v\: e^{-\vec{w}^2 v}\, {v}^{d/2-1} (1+v)^{-1}
\nonumber \\
&=& \frac{2^{d/2}}{\Gamma\bigl(1-\frac{d}{2}\bigr)\Gamma\bigl(\frac{d}{2}\bigr)}\bigl(y-1\bigr)^{-d/2}
\int_0^{\infty} \!\D v\: {v}^{d/2-1} (1+v)^{-1}  \left( v+\frac{y+1}{y-1}\right)^{-d/2}
\nonumber \\
&=& \frac{1}{\Gamma\bigl(1-\frac{d}{2}\bigr)\Gamma\bigl(\frac{d}{2}\bigr)}\left(\frac{2}{y-1}\right)^{d/2}
\int_0^1 \!\D u\: \bigl(1-u\bigr)^{d/2-1} \left( 1 + \frac{2}{y-1} u\right)^{-d/2}
\nonumber \\
&=& \frac{1}{\Gamma\bigl(1-\frac{d}{2}\bigr)\Gamma\bigl(\frac{d}{2}+1\bigr)}\left(\frac{2}{y-1}\right)^{d/2}
{}_2F_{1}\left(\frac{d}{2},1;\frac{d}{2}+1;-\frac{2}{y-1}\right)
\label{gl:2.17}
\EEA
where we first carried out the gaussian integration over $\vec{w}$ and then changed the integration variable.
By the further change $\mu=1-u$ we can go over from the combination $\frac{2}{y-1}$ to $\frac{2}{y+1}$ and finally obtain
\begin{subequations}\label{gl:voter-C2}
\begin{align}\label{gl:voter-C2-bas}
C(ys,s) &= \frac{1}{\Gamma\bigl(1-\frac{d}{2}\bigr)\Gamma\bigl(\frac{d}{2}\bigr)}\left(\frac{2}{y+1}\right)^{d/2}
\int_0^1 \!\D \mu\: \mu^{d/2-1} \left( 1 - \frac{2}{y+1} \mu\right)^{-d/2}
\nonumber \\
&= \frac{1}{\Gamma\bigl(1-\frac{d}{2}\bigr)\Gamma\bigl(\frac{d}{2}+1\bigr)}\left(\frac{2}{y+1}\right)^{d/2}
{}_2F_{1}\left(\frac{d}{2},\frac{d}{2};\frac{d}{2}+1;\frac{2}{y+1}\right)
\;\; ; \;\; d<2
\end{align}
\end{subequations}
where obviously the limit $y\to 1$ can now be taken and $C(s,s)=1$.
The relation between (\ref{gl:2.17}) and (\ref{gl:voter-C2-bas}) is a well-known hypergeometric identity \cite{Abra65,Prud3}.
If $d=1$ we recover the well-known expression for the Glauber-Ising chain, e.g. \cite{Godr00a,Lipp00,Corb24c,Henk25b}
\BEQ
C^{(1D)}(ys,s) = \frac{2}{\pi}\left(\frac{2}{y+1}\right)^{1/2} {}_2F_1\left(\demi,\demi;\frac{3}{2};\frac{2}{y+1}\right) = \frac{2}{\pi} \arcsin \sqrt{\frac{2}{y+1}\,}
\EEQ

{\bf 2.} In the second case $d>2$, we now have $b=d/2-1$. Reusing the integral representation \cite[(13.2.5)]{Abra65}, we have from (\ref{gl:2.16}) for $y>1$
\BEA
C(ys,s) &=&  \frac{(2/\pi)^{d/2} \mathfrak{C}_0}{2^{d-2}\Gamma\bigl(\frac{d}{2}-1\bigr)} s^{-d/2+1}\bigl(y-1\bigr)^{-d/2}
\int_{\mathbb{R}^d} \!\D\vec{w}\: e^{-\vec{w}^2\frac{y+1}{y-1}} \int_0^{\infty} \!\D v\: e^{-\vec{w}^2 v}\,  (1+v)^{d/2-2}
\nonumber \\
&=& \frac{\mathfrak{C}_0}{2^{d/2-2}\Gamma\bigl(\frac{d}{2}-1\bigr)} s^{-d/2+1} \bigl(y-1\bigr)^{-d/2}
\int_0^{\infty} \!\D v\:  (1+v)^{d/2-2}  \left( v+\frac{y+1}{y-1}\right)^{-d/2}
\nonumber \\
&=& \frac{\mathfrak{C}_0}{2^{d/2-2}\Gamma\bigl(\frac{d}{2}-1\bigr)} s^{-d/2+1} \bigl(y-1\bigr)^{-d/2}
\int_0^1 \!\D u\: \left(1+\frac{2}{y-1} u\right)^{-d/2}
\label{gl:2.19} \\
&=& \frac{\mathfrak{C}_0}{2^{d/2-2}\Gamma\bigl(\frac{d}{2}-1\bigr)} s^{-d/2+1} \bigl(y+1\bigr)^{-d/2}
\int_0^1 \!\D \mu\: \left(1-\frac{2}{y+1} \mu\right)^{-d/2}
\nonumber
\EEA
where the last-but-one line is analogous to the last-but-one line in (\ref{gl:2.17}). The final result is recognised from the last line and reads
\addtocounter{equation}{-3}   
\begin{subequations}
\addtocounter{equation}{1}    
\begin{align}
C(ys,s) &= \frac{\mathfrak{C}_0}{2^{d/2-2}\Gamma\bigl(\frac{d}{2}-1\bigr)} s^{-d/2+1} \bigl(y+1\bigr)^{-d/2}
{}_2F_{1}\left(\frac{d}{2},1;2;\frac{2}{y+1}\right) \;\; ;  \;\; d>2
\label{gl:voter-C2-haut}
\\
&= \frac{\mathfrak{C}_0}{2^{d/2-1}\Gamma\bigl(\frac{d}{2}\bigr)}  s^{-(d/2-1)}
\left[ \bigl(y-1\bigr)^{-(d/2-1)} - \bigl(y+1\bigr)^{-(d/2-1)} \right]
\label{gl:voter-C2-elem}
\end{align}
\end{subequations}
\addtocounter{equation}{2}   
which one may  even recast in terms of elementary functions for $y>1$.
This last form (\ref{gl:voter-C2-elem}) is identical to several mean-field results in non-equilibrium dynamics for $d>d^*$
in simple ferromagnets, interacting particle systems or in surface growth, see \cite[Table 4.2]{Henk10}.

Eqs.~(\ref{gl:voter-C2}) are our final result for the two-time auto-correlator, when $d\ne 2$.
When studying the behaviour near to $y\approx 1$, the alternative forms (\ref{gl:2.17},\ref{gl:2.19}) or (\ref{gl:voter-C2-elem}) are useful.
There is an obvious power-law decay for $y\gg 1$ with $\lambda=d$. For $y\to 1$, we find a saturation for $d<2$ but a power-law in $y-1$ for $d>2$.
In this way, we have the analytic confirmation of the statements we drew from a few examples in figures~\ref{fig3}c and~\ref{fig4}c.
We see again that for $d<2$, the auto-correlator saturates as one would have anticipated for a magnetic system undergoing phase-ordering kinetics while for $d>d^*=2$,
the scaling function reduces to the one of mean-field theory for non-equilibrium critical dynamics.

\subsection{Two-time response}


For deriving the equations of motion for the response function, we recall that the voter model can be viewed as a kinetic Ising model in simultaneous contact with two distinct baths:
one creating a Glauber single spin-flip dynamics at temperature $T=0$ and the other creating a Kawasaki spin-exchange dynamics at $T=\infty$ \cite{Droz89,Krap92}.
We introduce a magnetic perturbation into the Glauber part of the dynamics.
Then we can appeal to the derivation of response functions in the Glauber-Ising model, follow the lines of \cite{Godr00a} and derive an equation of
motion for the local order-parameter $\langle\sigma_{\vec{n}}\rangle(t)$ to leading order in an external magnetic field $h_{\vec{m}}(s)$. For the response function
$R(t,s;\vec{n}-\vec{m})=\left.\frac{\delta \langle \sigma_{\vec{n}}\rangle(t)}{\delta h_{\vec{m}}(s)}\right|_{h=0}$, in the continuum limit this leads to
\BEQ \label{gl:2.20}
\partial_t R(t,s;\vec{r}) = \demi \Delta_{\vec{r}} R(t,s;\vec{r}) \;\; , \;\;
R(s,s;\vec{r}) = \demi \bigl( 1 - C(s;\vec{2})\bigr)\delta(\vec{r})
\EEQ
where the equal-time response is related to the single-time correlator with a spatial distance of two lattice sites.
This must be worked out in the long-time limit $s\to\infty$ which we do in the non-trivial case when $d<2$
\BEA
\lefteqn{R(s,s;\vec{r}) \simeq \left(\demi -\demi \frac{1}{\Gamma\bigl(1-\frac{d}{2}\bigr)} e^{-1/s}\,
U\left(\frac{d}{2},\frac{d}{2};\frac{1}{s}\right)\right)\delta(\vec{r}) }
\nonumber \\
&=&\left( \demi -\frac{e^{-1/s}}{2\Gamma\bigl(1-\frac{d}{2}\bigr)}
\left[ \frac{\Gamma\bigl(\frac{d}{2}-1\bigr)}{\Gamma\bigl(\frac{d}{2}\bigr)} \left(\frac{1}{s}\right)^{1-d/2}
{}_1F_1\left(1;2-\frac{d}{2};\frac{1}{s}\right)
+\frac{\Gamma\bigl(1-\frac{d}{2}\bigr)}{\Gamma\bigl(1\bigr)} {}_1F_1\left(\frac{d}{2};\frac{d}{2};\frac{1}{s}\right) \right]\right)\delta(\vec{r})
\nonumber \\
&\simeq& \left(\demi + \frac{1}{(2-d)\Gamma\bigl(1-\frac{d}{2}\bigr)} s^{d/2-1}
\left( 1 + {\rm O}\bigl(\frac{1}{s}\bigr)\right) -\demi + {\rm O}\bigl(\frac{1}{s}\bigr)\right)\delta(\vec{r})
\nonumber \\
&=& \frac{s^{d/2-1}}{2\Gamma\big(2-\frac{d}{2}\bigr)} \delta(\vec{r}) + \ldots \;\; ; \;\; d<2
\label{gl:2.21}
\EEA
The important piece of information is the scaling of this as a function of the waiting time $s$.
The solution of the equation of motion (\ref{gl:2.20}) proceeds via straightforward Fourier transforms and reads (once more, $d$ becomes a continuous parameter)
\BEA
\lefteqn{ R(t,s;\vec{r}) = \frac{s^{d/2-1}}{2\bigl(2\pi\bigr)^d\Gamma\bigl(2-\frac{d}{2}\bigr)}
\int_{\mathbb{R}^d}\!\D\vec{q}\: e^{\II\vec{q}\cdot\vec{r}-\demi\vec{q}^2(t-s)}
\int_{\mathbb{R}^d} \!\D\vec{r}'\: e^{-\II \vec{q}\cdot\vec{r}'} \delta\bigl(\vec{r}'\,\bigr) }
\nonumber \\
&=& \frac{s^{d/2-1}}{2\bigl(2\pi\bigr)^d\Gamma\bigl(2-\frac{d}{2}\bigr)} \int_{\mathbb{R}^d}\!\D\vec{q}\: e^{\II\vec{q}\cdot\vec{r}-\demi\vec{q}^2(t-s)}
\:=\: \frac{s^{d/2-1}}{2\bigl(2\pi\bigr)^{d/2}\Gamma\bigl(2-\frac{d}{2}\bigr)} \bigl(t-s\bigr)^{-d/2}\, e^{-\demi \vec{r}^2/(t-s)} ~~~~~
\label{gl:2.22}
\EEA
so that the end result is
\begin{subequations} \label{gl:voter-R2}
\BEQ \label{gl:voter-R2-bas}
R(ys,s;\vec{r}) = \mathfrak{R}_0\, s^{-1} \bigl( y-1\bigr)^{-d/2} \exp\left[ - \demi \frac{\vec{r}^2}{s\bigl(y-1\bigr)} \right] \;\; ; \;\; d<2
\EEQ
where we need not specify the normalisation constant $\mathfrak{R}_0$.\footnote{Its value is important for a discussion of the fluctuation-dissipation ratio,
brilliantly done in \cite{Sast03,Oliv03,Corb24f}.} The expected exponent equality $\lambda_R=\lambda_C=d$ (\ref{lambda}) is confirmed.

For $d>2$, we can repeat the same calculation which has led to (\ref{gl:2.21}).
It is enough to notice the end result $R(s,s;\vec{r})=\mbox{\rm cste. } \delta(\vec{r}) + {\rm O}\bigl(s^{1-d/2}\bigr)$.
Then the same kind of straightforward calculation leads to
\BEQ \label{gl:voter-R2-haut}
R(ys,s;\vec{r}) = \mathfrak{R}_0 \,s^{-d/2}\, \bigl( y-1\bigr)^{-d/2} \exp\left[ - \demi \frac{\vec{r}^2}{s\bigl(y-1\bigr)} \right] \;\; ; \;\; d>2
\EEQ
\end{subequations}
which is the form found in mean-field theory in  non-equilibrium critical dynamics.

Eqs.~(\ref{gl:voter-R2}) are our end result for the two-time response.
We can use these expressions to reconfirm the three defining ageing properties (I,II,III) from section~\ref{sec:1}.
Then eqs.~(\ref{gl:voter-R2}) do agree with all scaling expectations (\ref{gl:local}).
We stress the gaussian form of the result, even if the voter model itself is not gaussian for $d<2$ as is shown by the form of its correlators.

\begin{table}[tb]
\begin{center}
\begin{tabular}{|c|lllll|}  \hline
       & ~$\mathpzc{z}$~ & ~$\lambda$~ & ~$b$~     & ~$a$~     & ~$a'$~ \\ \hline
$d<2$  &  ~$2$           &  ~$d$       & $0$       & $0$       & $d/2-1$ \\
$d=2$  &  ~$2$           &  ~$d$       & $0$ (log) & $0$ (log) & $d/2-1$ \\
$d>2$  &  ~$2$           &  ~$d$       & $d/2-1$   & $d/2-1$   & $d/2-1$ \\ \hline
\end{tabular}\end{center}
\caption[tab2]{Values of the exponents of non-equilibrium ageing (\ref{gl:local}) in the voter model. Logarithmic factors in $d=2$, if they occur, are indicated.}
\label{tab:2}
\end{table}

Summarising, we collect in table~\ref{tab:2} the values of the exponents which describe the ageing (\ref{gl:local})
of the voter model and add the exponent $a'$ defined in (\ref{2reponseR}) below.
The apparent simplicity of the values of $\mathpzc{z}$ and $\lambda$ might one lead to believe in a superficial simplicity of the model,
but the values of $b$ and $a$ clearly show this to be incorrect.
The value $b=0$ for $d<2$ is the same as in phase-ordering kinetics but the voter model is distinguished from those by its well-known lack of surface tension \cite{Ligg85,Tome01,Krap10}.
This leads to $a=b=0$ for $d<2$ and brings the model more closely into the range of non-equilibrium critical dynamics.
For $d>2$, the voter model exponents and the scaling functions are those of mean-field theory of
critical non-equilibrium dynamics, including the identity $a'=a$.
The non-triviality of the voter model resides in its correlation functions, to be discussed further in section~\ref{sec:4}.

\subsection{The case $d=2$} \label{subsec:2.4}

The scaling behaviour for the upper critical dimension $d=d^*=2$ has peculiarities \cite{Krap92,Sast03,Hase06b,Corb24b}.
In the continuum limit, the equation of motion for the single-time correlator
\BEQ
\partial_t C(t;r) = \Delta_{\vec{r}} C(t;r) = \left( \partial_r^2 + \frac{1}{r} \partial_r \right) C(t;r)
\EEQ
is now to be solved by a modified scaling ansatz
\BEQ
C(t;r) = \bigl( \ln t\bigr)^{-\beta} F_C(\mathfrak{u}) \;\; , \;\; \mathfrak{u} = r\, t^{-1/2}
\EEQ
where once more we let $t\to\infty$ and $r\to\infty$ with $\mathfrak{u}$ fixed and where the constant $\beta$ must be determined.
This  leads to the differential equation
\BEQ
F_C''(\mathfrak{u}) + \frac{1}{\mathfrak{u}}F_C'(\mathfrak{u}) +\demi \mathfrak{u} F_C'(\mathfrak{u}) = -\beta \frac{1}{\ln t} F_C(\mathfrak{u}) \to 0
\EEQ
in the long-time limit $t\to\infty$ and gives the same equation as (\ref{gl:2.6}) in the special case $b=0$.
However, for $d=2$ the usual scaling behaviour is broken by a logarithmic factor and in addition there exist additive logarithmic corrections.
We can carry over the general solution (\ref{gl:gen-solC1}) wherein the requirement that $C(t;r)$
should vanish more fast than any power for $r\to\infty$ fixes $\mathscr{C}_1=0$. We also set arbitrarily $\mathscr{C}_2=1$. This leads to
\BEQ
C(t;r) = \ln^{-\beta} t\cdot e^{-{r}^2/4t}\, U\left(1,1;\frac{{r}^2}{4t}\right) = \ln^{-\beta} t\cdot E_1\left(\frac{{r}^2}{4t}\right)
\EEQ
where $E_1(z)$ denotes the exponential integral \cite{Abra65}.
With the asymptotics $E_1(z) \simeq -C_E -\ln z + {\rm O}(z)$ where $C_E=0.577\ldots$ is Euler's constant, we find for $r\to 0$
\BEQ
C(t;{r}) \simeq \ln \frac{{r}^2}{4} \cdot \ln^{-\beta} t \cdot \ln t - \frac{C_E}{\ln^{\beta} t} + \ldots
\EEQ
such that the leading term becomes $t$-independent if $\beta=1$.
Then the final expression for the single-time correlator is (up to an unspecified normalisation)
\BEQ \label{gl:voter-C1-d2}
C(t;{r}) = \frac{1}{\ln t} e^{-{r}^2/4t}\, U\left( 1,1;\frac{{r}^2}{4t}\right) = \frac{1}{\ln t} E_1\left(\frac{{r}^2}{4t}\right) \;\; ; \;\; d=2
\EEQ
and reproduces the anterior calculations of \cite{Krap92} and \cite[(7,20)]{Corb24b}. For a numerical test, see \cite{Tart15}. This completes (\ref{gl:voter-C1}).

To find the two-time auto-correlator, we go back to (\ref{gl:2tC}) but must again work out $C(s;\vec{r})$. From (\ref{gl:voter-C1-d2}) we read off that
$C(s;\vec{r})=\frac{1}{\ln s}F_C\bigl(\frac{r^2}{4s}\bigr)$.
With the identity ${}_2F_1\bigl(1,1;2;z)=-z^{-1}\ln(1-z)$ \cite{Abra65}, repeating the calculation which gave (\ref{gl:voter-C2-haut}),
now for $d=2$, leads to, up to a normalisation factor \cite{Sast03,Oliv03}
\BEQ \label{gl:voter-C2-d2}
C(ys,s) = \frac{\mathfrak{C}_0}{\ln s} \bigl(y+1\bigr)^{-1} {}_2F_1\left(1,1;2;\frac{2}{y+1}\right)
        = \frac{2\mathfrak{C}_0}{\ln s} \frac{1}{y+1}\ln \frac{y-1}{y+1} \;\; ; \;\; d=2
\EEQ
and completes (\ref{gl:voter-C2}).
Although this does have a logarithmic pre-factor $\frac{1}{\ln s}$, for $y\gg 1$ there is no leading logarithmic contribution in the scaling function,
as is well-known for critical-point ferromagnets at their upper critical dimension $d=d^*$ \cite{Hase06,Ebbi08}.

The two-time response function is found from (\ref{gl:2.20}), where we need the initial correlator
$C(s;\vec{2})$. From (\ref{gl:voter-C1-d2}) we have $C(s;\vec{2})\sim \frac{1}{\ln s}$
and insertion into (\ref{gl:voter-R2-haut}) produces, up to normalisation
\BEQ \label{gl:voter-R2-d2}
R(ys,s;\vec{r}) = \mathfrak{R}_0 \,\frac{1}{s\ln s}\, \bigl( y-1\bigr)^{-1} \exp\left[ - \demi \frac{\vec{r}^2}{s\bigl(y-1\bigr)} \right] \;\; ; \;\; d=2
\EEQ
and completes (\ref{gl:voter-R2}).
Then the exponents in table~\ref{tab:2} are continuous at $d=2$. The usual scaling (\ref{gl:local}) is broken by a logarithmic factor,
but the scaling functions $F_{C,R}$ can be obtained by taking the
$d\to 2$ limit in the expressions derived before for $d\ne 2$.

\section{Predictions from dynamical symmetries} \label{sec:3}

The forthcoming comparison of the exact results of section~\ref{sec:2} with the dynamical Schr\"odinger symmetry, see section~\ref{sec:4},
will be based on four distinct ingredients which we shall now briefly review.
Although the equations of motion for correlators and responses in the $d$-dimensional voter model are linear,
the constraint $\sigma_{\vec{n}}^2=1$ leads to non-trivial field theories. This makes the use of a field-theoretic apparatus necessary.
For $d\ne 2$, the final predictions are (\ref{2reponse},\ref{gl:3.17},\ref{gl:3.18}).

\subsection{Non-equilibrium field-theory}

This work is inserted into the context of non-equilibrium continuum field-theory \cite{Domi76,Jans76,Jans92,Taeu14}.
Implicitly, we shall always restrict to non-conserved dynamics (`model A') of the order-parameter $\phi$.
The average of an observable $\mathscr{A}$ is found from the functional integral
\BEQ \label{dynft}
\bigl\langle \mathscr{A}\bigr\rangle = \int \mathscr{D}\phi\mathscr{D}\wit{\phi}\; \mathscr{A}[\phi]\, e^{-{\cal J}[\phi,\wit{\phi}]}
\EEQ
which already includes probability conservation \cite{Taeu14}. The Janssen-de Dominicis action reads
\begin{align} \label{actionJdD}
{\cal J}[\phi,\wit{\phi}] &= \int \!\D t\D\vec{r}\: \left( \wit{\phi} \left( \partial_t - \Delta_{\vec{r}} - V'[\phi]\right)\phi - T \wit{\phi}^2 \right)
\end{align}
with the interaction $V'[\phi]$, the spatial laplacian $\Delta_{\vec{r}}$ (with usual re-scalings), the response scaling operator $\wit{\phi}$ and also contains
a thermal white noise of temperature $T$. We need not consider here any contribution from a noisy initial configuration.
The {\em deterministic action} ${\cal J}_0[\phi,\wit{\phi}]  = \lim_{T\to 0} {\cal J}[\phi,\wit{\phi}]$ refers to temperature $T=0$.
We define {\em deterministic averages} $\bigl\langle \cdot \bigr\rangle_0$ as
\BEQ \label{detave}
\bigl\langle \mathscr{A}\bigr\rangle_0 = \int \mathscr{D}\phi\mathscr{D}\wit{\phi}\; \mathscr{A}[\phi]\, e^{-{\cal J}_0[\phi,\wit{\phi}]}
\EEQ
by replacing in (\ref{dynft}) the action ${\cal J}$ by its deterministic part ${\cal J}_0$.
{}From causality considerations \cite{Jans92,Cala05,Taeu14} or Local-Scale-Invariance ({\sc lsi}) \cite{Pico04}
one has the Barg\-man su\-per\-se\-lec\-tion rules
\BEQ \label{Bargman}
\left\langle \overbrace{~\phi \cdots \phi~}^{\mbox{\rm ~~$n$ times~~}}
             \overbrace{ ~\wit{\phi} \cdots \wit{\phi}~}^{\mbox{\rm ~~$m$ times~~}}\right\rangle_0 \sim \delta_{n,m}
\EEQ
for the deterministic averages. Only observables built from an equal number of order-parameters $\phi$
and conjugate response operators $\wit{\phi}$ can have non-vanishing deterministic averages.
Response functions as in (\ref{gl:1}) can now be found directly as deterministic averages.
However, correlators are obtained from three-point response functions \cite{Pico04,Henk10}
\BEQ \label{corrFT}
C(t,s;r)
= \bigl\langle \phi(t,\vec{r}+\vec{r}_0)\phi(s,\vec{r}_0) \bigr\rangle
= T \int_0^{\infty} \!\D u \int_{\mathbb{R}^d} \!\D\vec{R}\:
\left\langle \phi(t,\vec{r}+\vec{r}_0) \phi(s,\vec{r}_0) \wit{\phi^2}(u,\vec{R}) \right\rangle_0
\EEQ
Evaluating this, we shall see below that $\wit{\phi}_2 := \wit{\phi^2}$ must be treated as a composite scaling operator.

\subsection{Schr\"odinger invariance}

Calculations using (\ref{actionJdD},\ref{detave}) share many aspects of equilibrium calculations.
The deterministic action ${\cal J}_0$ has the Schr\"odinger group as a dynamical
symmetry.\footnote{Explicit examples include free fields \cite{Henk03a} and the $(1+1)D$ Calogero model \cite{Shim21}.}
The Lie algebra is spanned by $\langle X_{\pm 1,0}, Y_{\pm\frac{1}{2}}, M_0\rangle$ (for simplicity in a $1D$ notation) where
\BEA
X_n &=& - t^{n+1}\partial_t - \frac{n+1}{2} t^n r\partial_r - (n+1) \delta t^n -\frac{n(n+1)}{4} {\cal M} t^{n-1} r^2 \nonumber \\
Y_m &=& - t^{m+\frac{1}{2}} \partial_r - \left( m + \frac{1}{2}\right) {\cal M} t^{m-\frac{1}{2}} \\
M_n &=& - t^n {\cal M} \nonumber
\EEA
which close into the Schr\"odinger Lie algebra $\mathfrak{sch}(1)$.
This is the most large and finite-dimensional Lie algebra which sends each solution of $\bigl(2{\cal M}\partial_t - \partial_r^2\bigr)\phi=0$ into
another solution, as already known to Jacobi and to Lie, see \cite{Duva24} and references therein.
In addition, $\delta$ is the scaling dimension and ${\cal M}>0$ the (non-relativistic) mass of the equilibrium scaling operator $\phi$.
The form of the scaling generator $X_0$ implies the dynamical exponent $\mathpzc{z}=2$.
The requirement of Schr\"odinger-covariance constrains the form of $n$-point deterministic averages or response functions \cite{Henk94}.
Hence the Schr\"odinger-algebra generators act as super-operators on the scaling operators $\phi,\wit{\phi}$.
We shall need  the two-point response function ($\mathscr{R}_0$ is a normalisation constant)
\BEA
\lefteqn{\hspace{-2.5truecm}R(t_a,t_b;r) = \left\langle \phi_a(t_a,\vec{r}) \wit{\phi}_b(t_b,\vec{0})\right\rangle_0
= {\delta({\cal M}_a + {\cal M}_b)}\, \mathscr{R}_0\, \delta_{\delta_a,\wit{\delta}_b}\: \Theta(t_a-t_b)} \nonumber \\
&\times&  \bigl( t_a - t_b\bigr)^{-2\delta_a} \exp\left[ - \frac{{\cal M}_a}{2} \frac{\vec{r}^2}{t_a-t_b} \right]
\label{2points}
\EEA
such that response operators have negative masses {$\wit{\cal M}={\cal M}_b=-{\cal M}_a=-{\cal M}<0$};
and especially the three-point response
\BEA
\lefteqn{\hspace{-1.5truecm}\bigl\langle \phi_a(t_a,\vec{r}_a) \phi_b(t_b,\vec{r}_b)\wit{\phi}_c(t_c,\vec{r}_c)\bigr\rangle_0 =
{\delta({\cal M}_a + {\cal M}_b + {\cal M}_c)} \Theta(t_{ac}) \Theta(t_{bc})\: }
\nonumber \\
&\times& t_{ac}^{-\delta_{ac,b}} t_{bc}^{-\delta_{bc,a}} t_{ab}^{-\delta_{ab,c}} \:
\exp\left[ -\frac{{\cal M}_a}{2} \frac{\vec{r}_{ac}^2}{t_{ac}} - \frac{{\cal M}_b}{2} \frac{\vec{r}_{bc}^2}{t_{bc}}\right]
\Phi_{ab,c}\left( \frac{\bigl[ \vec{r}_{ac}^2 t_{bc} - \vec{r}_{bc}^2 t_{ac} \bigr]^2}{t_{ab} t_{ac} t_{bc} }\right)
\label{3points}
\EEA
where we used that from (\ref{2points}) it follows $\wit{\delta}_c=\delta_c$ and with the shorthands
\BD
t_{ij} = t_i - t_j \;\; , \;\; \vec{r}_{ij} = \vec{r}_i - \vec{r}_j \;\; , \;\; \delta_{ij,k} = \delta_i + \delta_j - \delta_k
\ED
{Mass conservation  is only possible if the scaling operator $\phi_c$ is really a response operator $\wit{\phi}_c$
with a mass ${\cal M}_c=-({\cal M}_a + {\cal M}_b)$}.
The Heaviside functions $\Theta$ take the causality constraints in (\ref{2points},\ref{3points}) into account \cite{Henk03a}.
The function $\Phi_{ab,c}$ is not determined by Schr\"odinger-invariance.

In the applications below, we shall take into account that $\wit{\phi}_2:=\wit{\phi^2}$
must be considered as a composite scaling operator, with a non-trivial value
of its scaling dimension \cite{Moro11,Henr23,Anti25,Paga25}.

\subsection{Far from equilibrium observables} \label{subsec:3.3}

Physical ageing occurs far from equilibrium and is not time-translation-invariant.
Rather than dropping the time-translation generator $X_{-1}$ from the Schr\"odinger algebra, as for example explored in \cite{Henk06,Mini12},
we use an inspiration from dynamical symmetries of non-equilibrium systems \cite{Stoi22}
and propose to achieve far-from-equilibrium physics, without standard time-translation-invariance, by the following

\noindent
{\bf Postulate:} \cite{Henk25c} {\it The Lie algebra generator $X_n^{\rm equi}$ of a time-space symmetry of an equilibrium system becomes a symmetry
out-of-equilibrium by the change of representation}
\BEQ \label{gl:hyp}
X_n^{\rm equi} \mapsto X_n = e^{W(t)}\, X_n^{\rm equi}\, e^{-W(t)} \;\; , \;\; W(t) = \xi \ln t
\EEQ
{\it where $\xi$ is a dimensionless parameter whose value contributes to characterise the scaling operator $\phi$ on which $X_n$ acts.}

Let us consider two explicit examples: in the dilatation generator $X_0$
this merely leads to a modified scaling dimension $\delta_{\rm eff} = \delta-\xi$, since
\begin{subequations} \label{gl:Xgen}
\BEQ \label{gl:X0gen}
X_0^{\rm equi} \mapsto X_0 = -t\partial_t - \demi r\partial_r - \bigl(\delta - \xi\bigr)
\EEQ
and the time-translation generator $X_{-1}^{\rm equi}=-\partial_t$  turns into
\BEQ \label{gl:X-1gen}
X_{-1}^{\rm equi} \mapsto X_{-1} = -\partial_t + \frac{\xi}{t}
\EEQ
\end{subequations}
Significantly, in this new representation the scaling operators become
$\Phi(t) = t^{\xi} \phi(t) = e^{\xi \ln t}\phi(t)$ and the equilibrium response functions
(\ref{2points},\ref{3points}) are adapted into non-equilibrium ones via
(spatial arguments are suppressed for clarity)\footnote{This depends on the applicability of (\ref{gl:hyp}). Different changes of representation are conceivable \cite{Henk25c}.}
\BEA
\bigl\langle \phi_a(t_a)\wit{\phi}_b(t_b)\bigr\rangle_0            &\mapsto & t_a^{\xi_a} t_b^{\wit{\xi}_b} \: \bigl\langle \phi_a(t_a)\wit{\phi}_b(t_b)\bigr\rangle_0 \nonumber \\
\bigl\langle \phi_a(t_a)\phi_b(t_b)\wit{\phi}_c(t_c)\bigr\rangle_0 &\mapsto & t_a^{\xi_a} t_b^{\xi_b} t_c^{\wit{\xi}_c} \: \bigl\langle \phi_a(t_a) \phi_b(t_b)\wit{\phi}_c(t_c)\bigr\rangle_0
\label{reponseHE}
\EEA
This means that we now characterise a scaling operator $\phi$ by a pair of scaling dimensions $(\delta,\xi)$ and a response operator
$\wit{\phi}$ by a pair $(\wit{\delta},\wit{\xi})$.
As a consequence of Schr\"odinger-invariance (\ref{2points}), and the Bargman rule (\ref{Bargman}) with $n=m=1$, we have
\BEQ
\delta = \wit{\delta}
\EEQ
but $\xi$ and $\wit{\xi}$ remain independent.

\subsection{Irrelevance of non-linearities at long times} \label{subsec:3.4}

For completeness, we briefly mention another consequence of the change of representation (\ref{gl:hyp}),
although we shall not need it in the voter model since therein all equations of motion for correlators and responses are linear.

For non-equilibrium dynamics in models with an inversion-symmetry $\phi(t,\vec{r})\mapsto -\phi(t,\vec{r})$,
one generally expects for sufficiently long times an effective equation of motion \cite{Taeu14}
\BEQ \label{gl:phi-eff}
\left( \partial_t - \frac{1}{2{\cal M}} \Delta_{\vec{r}} \right) \phi(t,\vec{r}) = g \phi^3(t,\vec{r})
\EEQ
with a coupling $g$.
On the left-hand-side of (\ref{gl:phi-eff}) we have the Schr\"odinger operator $\mathscr{S}^{\rm equi}=\partial_t - \frac{1}{2{\cal M}}\Delta_{\vec{r}}$, which in the
new representation according to (\ref{gl:hyp}) will become
\BEQ \label{gl:phi-eff2}
{\mathscr{S}} = e^{\xi \ln t} \mathscr{S}^{\rm equi} e^{-\xi \ln t} = \partial_t - \frac{\xi}{t} - \frac{1}{2{\cal M}} \Delta_{\vec{r}}
\EEQ
and contains an additional $1/t$-potential. The original equation $\mathscr{S}^{\rm equi} \phi =g  \phi^3$ then becomes, with the notation of subsection~\ref{subsec:3.3}
\BEQ \label{gl:phi-eff3}
\left( t^{\xi}\, \mathscr{S}^{\rm equi}\, t^{-\xi} \right) \left( t^{\xi}\, \phi \right) = g\,t^{\xi} \left( t^{-\xi}\, {\Phi}\, \right)^{3}
~~\Longrightarrow~~
{\mathscr{S}}\,{\Phi} = \left( \partial_t - \frac{\xi}{t} - \frac{1}{2{\cal M}} \Delta_{\vec{r}} \right) {\Phi} = g\, t^{-2\xi}\, {\Phi}^{\,3}
\EEQ
This suggests that for long-times, the $1/t$-potential in (\ref{gl:phi-eff3})
should dominate over against any non-linear term, when $2\xi>1$ (higher-order power are even less relevant).
In addition, it must be taken into account that the coupling $g$ has a non-trivial dimension \cite{Stoi05} and this permits to circumvent the condition $\delta=\frac{d}{4}$ \cite{Henk25c}.

This might become important in the future in models with non-linear equations of motion, beyond the voter model under consideration in this work.

\subsection{Summary}

In physical ageing, observables include both response and correlation functions. The general line of argument is as follows.
First, response functions will be assumed to transform covariantly under Schr\"odinger-transformations.
Second, due to the relation (\ref{reponseHE}) which maps the responses {\em at} equilibrium to their non-equilibrium analogues, it can be
shown that in the effective equation of motion the eventual non-linear terms (if they occur)
are irrelevant for the leading long-time behaviour, if the auto-correlation exponent obeys
the criterion $2\xi>1$ \cite{Henk25c}.
Third, for the voter model we require non-equilibrium dynamics with $\mathpzc{z}=2$,
such that Schr\"odinger-invariance in its non-equilibrium representation is indeed the relevant dynamical symmetry.
Combining (\ref{reponseHE},\ref{2points}) for the two-time response \cite{Henk06,Henk25c} we have for $t>s$
\begin{subequations} \label{2reponse}
\begin{align} \label{2reponseR}
R(t,s;r) = \left\langle {\phi}(t,\vec{r}) {{\wit{\phi}}}(s,\vec{0})\right\rangle
= \mathscr{R}_0 \, s^{-1-a} \left( \frac{t}{s}\right)^{1+a'-\lambda_R/2}
\left(\frac{t}{s} -1 \right)^{-1-a'} \exp\left[ -\frac{\cal M}{2} \frac{r^2}{t-s} \right]
\end{align}
where the exponents $a,a',\lambda_R$ of ageing are related to $\delta,\xi,\wit{\xi}$ as follows
\BEQ \label{2reponse-exp}
\frac{\lambda_R}{2} = 2\delta -\xi \;\; , \;\; 1+a = 2\delta -\xi - \wit{\xi} \;\; , \;\; a'-a = \xi +\wit{\xi}
\EEQ
\end{subequations}
and ${\cal M},\mathscr{R}_0$ are non-universal, dimensionful constants.

The single-time correlator can be shown to take the form \cite{Henk25d}
\BEA \label{gl:3.17}
C(t;\vec{r}) = \left\langle \phi(t,\vec{r}) \phi(t,\vec{0}) \right\rangle
= t^{-b} e^{-\frac{\cal M}{4}\frac{\vec{r}^2}{t}}\, U\left( 2\wit{\xi}_2+1, 4\wit{\delta}_2-\frac{d}{2};\frac{\cal M}{4}\frac{\vec{r}^2}{t}\right)
\EEA
up to an overall normalisation and where $U$ is a Kummer/Tricomi confluent hypergeometric function \cite{Abra65} and $b$ is given below in (\ref{gl:3.18b}).
This functional form depends only on the properties of the composite response field
$\wit{\phi}_2=\wit{\phi^2}$, which has the pair of scaling dimensions $(2\wit{\delta}_2,2\wit{\xi}_2)$.
In addition, there is the constraint $\wit{\delta}_2=\wit{\delta}=\delta$ \cite{Henk25d}.
In comparison with models, the parameters $\wit{\delta}_2$, $\wit{\xi}_2$ and $b$ as well as the non-universal constant $\cal M$ must be fixed.

The two-time auto-correlator has the form, again up to an overall normalisation \cite{Henk25d}
\begin{subequations} \label{gl:3.18}
\begin{align} \label{gl:3.18a}
C(ys,s) &= \left\langle \phi(ys,\vec{0}) \phi(s,\vec{0}) \right\rangle
       \:=\: s^{-b}\, y^{\xi+d/2-2\wit{\delta}_2} \big(y+1\bigr)^{-d/2-\nu} \bigl(y-1\bigr)^{-2(\delta-\wit{\delta}_2)+\nu} \nonumber \\
&  ~~\times F_1\left( 2\wit{\xi}_2+1,2\wit{\delta}_2-d/2,d/2+\nu;2+d/2+2\wit{\xi}_2-2\wit{\delta}_2;\frac{1}{y},\frac{2}{y+1}\right)
\end{align}
where $F_1$ is an Appell hypergeometric function \cite{Prud3}. As for the single-time correlator, this functional form depends only on properties of the
composite response field $\wit{\phi}_2 =\wit{\phi^2}$. Below, for comparison with the voter model,
we shall merely need the identity $F_1\bigl(a,0,b;c;w,z) = {}_2F_1\bigl(a,b;c;z)$.
The simplified form (\ref{gl:3.18a}), which will turn out to be sufficient for our purposes, depends on the additional hypothesis that
$\Psi(Y) = \int_{\mathbb{R}^d} \!\D\vec{P}\: e^{-\demi{\cal M}\vec{P}^2}\Phi_{\phi\phi,\wit{\phi}_2}\bigl(\vec{P}^2 Y\bigr) \stackrel{!}{=} \Psi_{\infty} Y^{\nu}$
becomes a simple power where $\Phi_{\phi\phi,\wit{\phi}_2}$ is a three-point function not fixed by Schr\"odinger-invariance \cite{Henk25d}.
It will turn out to hold true for the voter model. For $y\gg 1$ we can immediately read off
\BEQ \label{gl:3.18b}
\frac{\lambda_C}{2} = 2\delta - \xi \;\; , \;\; b = 4\wit{\delta}_2 -2\xi -2\wit{\xi}_2 -1 - \frac{d}{2}
\EEQ
\end{subequations}
and comparison with (\ref{2reponse-exp}) shows that indeed $\lambda_C=\lambda_R$, as expected from (\ref{lambda}).

In comparison with models, the parameters $\wit{\delta}_2$, $\wit{\xi}_2$, $\wit{\xi}$, $\xi$ as well as $\nu$ and the non-universal constant $\cal M$ must be fixed consistently
for the three observables (\ref{2reponseR},\ref{gl:3.17},\ref{gl:3.18a}) and with the parameters (\ref{2reponse-exp},\ref{gl:3.18b}).
It is remarkably that under the assumptions made, these explicit predictions merely depend on the hypothesis of a specific dynamical symmetry.
Universal information on a specific model can only enter through the values of the finite parameter set $(\wit{\delta}_2,\wit{\xi}_2,\wit{\xi},\xi)$.

\section{Comparison with the voter model} \label{sec:4}

At long last, we can confront the explicit voter model results of section~\ref{sec:2} with the predictions of Schr\"odinger-invariance from section~\ref{sec:3}.
This constitutes the central aspect of this work. The physical background of the voter model makes it appear appropriate to consider that the source of the noise is
in the reservoir such that a setting as non-equilibrium critical dynamics is adequate. We shall consider first the case where $d\ne 2$ and shall return to the
marginal case $d=2$ at the end, since it will require a separate discussion.
The final results for the various scaling dimensions $\wit{\delta}_2,\wit{\xi}_2,\wit{\xi},\xi$ are gathered in table~\ref{tab:1} where
for comparison we also include the the values obtained \cite{Henk25d} in the $1D$ Glauber-Ising model quenched to $T=0$ and the $d$-dimensional spherical model quenched onto $T=T_c$.

\subsection{Two-time response}

The two-time response functions is studied first since its form is independent of any noise at all.
Comparing (\ref{gl:voter-R2}) with the Schr\"odinger-invariance prediction (\ref{2reponse}),
a general agreement of the scaling functions is apparent. It is then possible to identify the parameters. In our normalisation, we find
\begin{subequations} \label{gl:compR}
\begin{align}
d<2 &~:~~ 1+a = 1\;\; \hspace{0.1cm}, \;\; 1+a' = \frac{d}{2} \;\; , \;\; 1+a'-\frac{\lambda}{2} = 0 \;\; , \;\; {\cal M}=1\\
d>2 &~:~~ 1+a = \frac{d}{2} \;\;    , \;\; 1+a' = \frac{d}{2} \;\; , \;\; 1+a'-\frac{\lambda}{2} = 0 \;\; , \;\; {\cal M}=1
\end{align}
\end{subequations}
This gives some of the entries in table~\ref{tab:2}.
Using the relationship (\ref{2reponse-exp}) with the scaling dimensions, the corresponding entries in table~\ref{tab:1} are produced.

\subsection{Single-time correlator}

Next, we consider the single-time correlator. Again, consider first the generic voter model scaling from
(\ref{gl:voter-C1}), especially as expressed through the Kummer function $U$ and then
compare with Schr\"odinger-invariance prediction (\ref{gl:3.17}).
The general agreement already confirms that our `hypoth\`ese de travail', namely that the voter model belongs to the
systems undergoing non-equilibrium critical dynamics, is justified.
It is only the properties of the composite response operator $\wit{\phi}_2$,
namely  the scaling dimensions $(2\wit{\delta}_2,2\wit{\xi}_2)$, which determine the form of the scaling function.
We can identify
\begin{subequations} \label{gl:compC1}
\begin{align}
d<2 &~:~~ b=0\;\; \hspace{0.7cm},     \;\; 2\wit{\xi}_2+1 = \frac{d}{2} \;\; , \;\; 4\wit{\delta}_2 - \frac{d}{2} = \frac{d}{2} \;\; , \;\; {\cal M}=1 \\
d>2 &~:~~ b = \frac{d}{2} - 1  \;\; , \;\; 2\wit{\xi}_2+1 = 1           \;\; , \;\; 4\wit{\delta}_2 - \frac{d}{2} = \frac{d}{2} \;\; , \;\; {\cal M}=1
\end{align}
\end{subequations}
The ageing exponent is listed in table~\ref{tab:2} and are fully consistent with the values read off from the response before,
while the other two scaling dimensions are included in table~\ref{tab:1}. There is total agreement.
For completeness, we also list the value of the non-universal mass $\cal M$,
which is the same as for the time-space response (\ref{gl:compR}), as it should be.

\subsection{Two-time auto-correlator}

\begin{table}
\begin{center}
\begin{tabular}{|lrr|ccccc|}  \hline
\multicolumn{3}{|l|}{~} & \multicolumn{5}{r|}{~} \\[-0.4cm] 
\multicolumn{3}{|l|}{model}              & ~$\delta=\wit{\delta}$~ & ~$\xi$~  & ~$\wit{\xi}$~ & ~$\wit{\delta}_2$~ & ~$\wit{\xi}_2$~ \\ \hline
Glauber-Ising & ~$T=0$~~    & ~$d=1$~    & $1/4$                   & $0$      & $-1/2$        & $1/4$              & $-1/4$          \\[0.2cm]
voter         &             & ~$0<d<2$~  & $d/4$                   & $0$      & $d/2-1$       & $d/4$              & $\demi\left(\frac{d}{2}-1\right)$ \\
voter         &             & ~$2<d$~    & $d/4$                   & $0$      & $0$           & $d/4$              & $0$             \\[0.2cm]
spherical     & ~$T=T_c$~   & ~$2<d<4$~  & $d/4$                   & $1-d/4$  & $d/4-1$       & $d/4$              & $d/4-1$         \\
spherical     & ~$T=T_c$~   & ~$4<d$~    & $d/4$                   & $0$      & $0$           & $d/4$              & $0$             \\ \hline
\end{tabular}\end{center}
\caption[tab1]{Values of the scaling dimensions needed in the calculation of correlators and responses in exactly solved models.
\label{tab:1}
}
\end{table}

We continue with the two-time correlator. The specific results of the voter model are in (\ref{gl:voter-C2}) and should be compared with the prediction
(\ref{gl:3.18}) of Schr\"odinger-invariance. Here, it is important to notice that the single-time correlator has already yielded that $2\wit{\delta}_2=\frac{d}{2}$, such
that the second parameter in the Appell function $F_1$ actually vanishes. We can then use the identity \cite{Prud3}
\BEQ
F_1\bigl(a,0,b;c;w,z\bigr) = {}_2F_1\bigl(a,b;c;z) = {}_2F_1\bigl(b,a;c;z)
\EEQ
which allows us to simplify the prediction (\ref{gl:3.18}) as follows
\BEQ \label{gl:4:SchC2}
C(ys,s) = s^{-b} y^{\xi} \bigl(y+1\bigr)^{-d/2-\nu} \bigl( y-1\bigr)^{\nu}\,
{}_2F_1\left( 2\wit{\xi}_2+1, \frac{d}{2}+\nu;2\wit{\xi}_2+2;\frac{2}{y+1}\right) \;\; ; \;\; \wit{\delta}_2=\frac{d}{4}
\EEQ
Then the same kind of qualitative comparison of the voter model results (\ref{gl:voter-C2})
with the Schr\"odinger-invariance prediction (\ref{gl:4:SchC2}) becomes possible
and we observe an overall agreement. It is now possible to identify the parameters
\begin{subequations} \label{gl:compC2}
\begin{align}
d<2 &~:~~ b=0\;\; \hspace{0.7cm},     \;\; 2\wit{\xi}_2+1 = \frac{d}{2} \;\; , \;\; \wit{\delta}_2  = \frac{d}{4} \;\; , \;\; \nu = 0  \\
d>2 &~:~~ b = \frac{d}{2} - 1  \;\; , \;\; 2\wit{\xi}_2+1 = 1           \;\; , \;\; \wit{\delta}_2  = \frac{d}{4} \;\; , \;\; \nu =  0
\end{align}
\end{subequations}
and we also recall from the response function that we had already fixed $\xi=0$.
Remarkably, the results of (\ref{gl:compC2}), from the two-time correlator,
fully agree with the results of the more early comparison (\ref{gl:compC1}) based on the single-time correlator.
Fixing $\nu=0$ means that the unknown three-point function $\Psi$
from Schr\"odinger covariance of the three-point response $\bigl\langle\phi\phi\wit{\phi^2}\bigr\rangle$
really takes the most simple form possible, namely a constant.
For $d<2$ it can be obtained from the condition that $C(ys,s)$ should saturate as $y\to 1^+$.
These three successful comparisons mean that the entire set of observables,
either from the scaling functions of the two-time response, the single-time time-space correlator or the two-time auto-correlator, is consistently
described by the consequences of the  dynamical Schr\"odinger symmetry of the voter model.
After those examples treated in \cite{Henk25d}, the voter model is only the third known example where this dynamical symmetry has been confirmed so extensively.
To obtain this confirmation was the aim of this work.

Table~\ref{tab:1} shows that the form of the scaling functions depends essentially on the parameters $\wit{\xi}$ and $\wit{\xi}_2$
which describe the change of representation of the
response operator $\wit{\phi}$ and its composite $\wit{\phi}_2$ which are most removed from any specific property of the lattice model and which owe their existence to
Janssen-de Dominicis theory.\footnote{Since for $d<2$, $2\wit{\xi}\ne 2\wit{\xi}_2$,
the response operator $\wit{\phi}_2$ is non-trivial and $\wit{\phi}_2=\wit{\phi^2}\ne \bigl(\wit{\phi}\,\bigr)^2$.}
This operator must be involved in the coupling between the system and the heat bath.

In table~\ref{tab:1} we also observe that in the mean-field regime for $d>d^*$, the various scaling dimensions are the same for the voter model and the critical spherical model.
However, the values of $d^*$ are distinct in the two models. Both are described by the same (gaussian) mean-field theory for $d>4$, but for $2<d<4$ the voter model has a mean-field
regime without a spherical model analogue. In the fluctuation-dominated cases $d<d^*$,
both models have different values of their scaling dimensions and hence are in distinct universality classes.

\subsection{The case $d=2$}


The ubiquitous logarithmic factors in the $2D$ voter model impede the use of the non-equilibrium representation (\ref{gl:hyp}) generated by the standard choice $W(t)=\xi\ln t$.
But such logarithmic factors may be obtained if we  consider instead the representation
\BEQ \label{gl:nr}
W(t) = \Xi \ln \bigl( \ln t\bigr)
\EEQ
In particular, in comparison with the standard representation, this implies $\xi=\wit{\xi}=\wit{\xi}_2=0$. The map (\ref{reponseHE}) must be modified accordingly.
The predictions of Schr\"odinger-invariance from section~\ref{sec:3} are then recast as follows, for quasi-primary scaling operators. The two-time response function becomes
\BEQ \label{gl:SchLR2}
R(t,s;\vec{r}) = \bigl(\ln t\bigr)^{\Xi} \bigl(\ln s\bigr)^{\wit{\Xi}} \bigl( t-s\bigr)^{-2\delta} \exp\left[ - \frac{\cal M}{2} \frac{\vec{r}^2}{t-s}\right]
\;\; ; \;\; \delta = \wit{\delta} \;\; , \;\; \wit{\cal M} = - {\cal M}
\EEQ
with the new parameters $\Xi$ and $\wit{\Xi}$ which characterise the representation of the scaling operators $\phi$ and $\wit{\phi}$. The usual constraints are also listed.
The single-time correlator now reads, see \cite{Henk25d}
\BEQ
C(t;r) = \bigl( \ln t\bigr)^{2\Xi}\, \int_0^1 \!\D v\: \left( \ln \bigl(1-v\bigr) \right)^{2\wit{\Xi}_2} v^{d/2-4\wit{\delta}_2} \exp\left[-\frac{\cal M}{4}\frac{r^2}{t}\frac{1}{v}\right]
\EEQ
and contains the new parameters $\Xi$ and $\wit{\Xi}_2$. We recall that there is the constraint $\wit{\delta}_2=\wit{\delta}=\delta$ \cite{Henk25d}.
Only if it happens that $\wit{\Xi}_2=0$, the previous simplifications can be reused and we find the simplified expression
\BEQ \label{gl:SchLC1}
C(t;r) = \bigl( \ln t\bigr)^{2\Xi}\,
e^{-\frac{\cal M}{4} \frac{r^2}{t}}\, U\left( 1, 4\wit{\delta}_2-\frac{d}{2};\frac{\cal M}{4} \frac{r^2}{t}\right) \;\; ; \;\; \wit{\Xi}_2=0
\EEQ
Finally, the two-time auto-correlator becomes, see \cite{Henk25d}
\BEA
\lefteqn{C(ys,s) = \bigl(\ln ys\bigr)^{\Xi} \bigl(\ln s\bigr)^{{\Xi}} y^{d/2-2\wit{\delta}_2} \bigl(y-1\bigr)^{-2(\delta-\wit{\delta}_2)} \bigl(y+1\bigr)^{-d/2}}   \\
& & \times \int_0^1 \!\D w\: \left( \ln w\right)^{2\wit{\Xi}_2} \bigl(1-w\bigr)^{d/2-2\wit{\delta}_2} \bigl(1-y^{-1}w\bigr)^{d/2-2\wit{\delta}_2} \bigl(1-\frac{2}{y+1}w\bigr)^{-d/2}
\Psi\left(\frac{y-1}{y+1-2w}\right)
\nonumber
\EEA
Once more, under the simplifying circumstance that $\wit{\Xi}_2=0$, and if the ansatz $\Psi(Y)\stackrel{!}{=}\Psi_{\infty} Y^{\nu}$ holds, we find the simplification
\begin{align} \label{gl:SchLC2}
C(ys,s) &= \bigl(\ln ys\bigr)^{\Xi} \bigl(\ln s\bigr)^{{\Xi}} y^{d/2-2\wit{\delta}_2} \big(y+1\bigr)^{-d/2-\nu} \bigl(y-1\bigr)^{-2(\delta-\wit{\delta}_2)+\nu} \nonumber \\
&  ~~\times F_1\left( 1,2\wit{\delta}_2-\frac{d}{2},\frac{d}{2}+\nu;2+\frac{d}{2}-2\wit{\delta}_2;\frac{1}{y},\frac{2}{y+1}\right)
\;\; ; \;\; \wit{\Xi}_2=0
\end{align}
In the predictions (\ref{gl:SchLR2},\ref{gl:SchLC1},\ref{gl:SchLC2}), non-universal normalisation factors have been suppressed.

These predictions are now compared with the results of the $2D$ voter model for the two-time response (\ref{gl:voter-R2-d2}), the single-time correlator (\ref{gl:voter-C1-d2}) and the
two-time correlator (\ref{gl:voter-C2-d2}), respectively.

Herein, the values of the non-logarithmic parameters can be read off from table~\ref{tab:1} by taking the limit $d\to 2$.
It will turn out that indeed the simplification $\wit{\Xi}_2=0$ holds,
such that we can compare the $2D$ voter model results with the simplified predictions. From the response function,
the new information merely concerns the logarithmic prefactors. Since $\ln ys = \ln s + \ln y \simeq \ln s + \ldots$ for large times, we have the
additional condition $\Xi + \wit{\Xi}=-1$ and $2\delta=2\wit{\delta}=1$, as well as ${\cal M}=1$ in the chosen normalisation.
Similarly, the single-time correlator is modified by a logarithmic factor. This leads to $2\Xi =-1$. On the other hand, the
form of the scaling function of $C(t;r)$ gives once more $2\wit{\delta}_2=\frac{d}{2}=1$ and reproduces ${\cal M}=1$, as it should.
Finally, for the two-time correlator, the logarithmic factors lead again to the condition $2\Xi =-1$.
As before, we can then further simplify the Schr\"odinger-invariance prediction (\ref{gl:SchLC2}) and obtain
\BEQ
C(ys,s) = \left( \ln s\right)^{-1} \bigl(y+1\bigr)^{-d/2-\nu} \bigl( y-1\bigr)^{\nu}\, {}_2F_1\left( 1, \frac{d}{2}+\nu; 2;\frac{2}{y+1}\right)
\;\; ; \;\; \wit{\Xi}_2=0 \;\; , \;\; \wit{\delta}_2 =\frac{d}{4}
\EEQ
and the agreement with the $2D$ voter model is achieved for $\nu=0$.
In conclusion, the new non-standard change of representation (\ref{gl:nr}) does reproduce the observables in the $2D$ voter model
with the parameters
\BEQ \label{gl:4.13}
\Xi = \wit{\Xi} = -\demi \;\; , \;\; \wit{\Xi}_2 = 0 \;\; , \;\; \delta=\wit{\delta}=\wit{\delta}_2 = \demi \;\; , \;\; \nu=0
\EEQ
Again, we see that the response scaling operator $\wit{\phi}_2$ is non-trivial, since $\wit{\Xi}\ne \wit{\Xi}_2$.
This is the first physical example of a non-standard choice of representation $W(t)$.

\section{Conclusions} \label{sec:5}

Studying the voter model in the continuum limit, for any spatial dimension $d>0$, has led us to revisit several aspects of this much-studied system.
We considered it here as a paradigmatic example of non-equilibrium scaling and ageing behaviour for models which need not necessarily obey the requirement of
detailed balance. In particular, we obtained explicit expressions for the scaling functions of single-time and two-time correlators and two-time responses, given
for $d\ne 2$ in (\ref{gl:voter-C1},\ref{gl:voter-C2}) and (\ref{gl:voter-R2}). These already
suggest that the physics of the voter model is in many {aspects} more close to the one of non-equilibrium critical dynamics than to phase-ordering kinetics, although
it shares with the latter one the property that the interior of ordered clusters is completely frozen and cannot be broken up by a thermal activation. Phase-ordering
kinetics is driven by the interface tension between domains \cite{Bray94a}
and is not possible in the voter model without an interface tension. The evolution of voter model's microscopic clusters can
only proceed via surface diffusion at the boundaries of these clusters \cite{Dorn01}.

This physical point of view is re-enforced by the successful comparison of the form of all these scaling functions with the predictions of Schr\"odinger-invariance since
it was explicitly assumed in these comparisons that the source of the noise comes from the thermal bath, as it is the case in non-equilibrium critical dynamics.
The physical basis of this is the assumption of covariance of the two-point and three-point response functions $\langle \phi\wit{\phi}\rangle$ and $\langle \phi\phi\wit{\phi^2}\rangle$
under Schr\"odinger-transformations, to be followed by mapping these results to a non-equilibrium representation.
We have seen explicitly that this comparison requires the use of two distinct non-equilibrium representations of the Schr\"odinger algebra:
a standard one for all dimensions $d\ne 2$, see table~\ref{tab:1}, and a new specific one for the marginal case $d=2$, see eq.~(\ref{gl:4.13}).
The voter model thereby becomes an important example for the applicability of a dynamical symmetry for the
explicit and model-independent calculation of the universal form of scaling functions in physical ageing.
Table~\ref{tab:1} also shows that the two models solved for all $d>0$, namely the spherical and the voter model, deviate from mean-field theory, for $d<d^*$, in distinct ways.
It would be of interest if the explicit predictions of the
scaling functions in eqs.~(\ref{2reponseR},\ref{gl:3.17},\ref{gl:3.18a}), valid in systems with dynamical exponent $\mathpzc{z}=2$ and undergoing non-equilibrium critical dynamics,
could be tested in other (quantum) systems \cite{Giam16}.
Just why the shape of the scaling functions should be essentially fixed by the composite response operator $\wit{\phi^2}$ is left as an open question.

Further extensions, notably towards phase-ordering kinetics which naturally have $\mathpzc{z}=2$, are under active investigation \cite{Henk25e}.

\noindent
{\bf Acknowledgements:}
{This work was supported by the french ANR-PRME UNIOPEN (ANR-22-CE30-0004-01), PHC RILA (Dossier 51305UC)
and Bulgarian National Science Fund, grant KP-06-N88/3}.


\end{document}